\documentclass[11pt]{article}
\pdfoutput=1
\usepackage{jheppub}
\usepackage{subcaption}
\usepackage{tikz}

\makeatletter

\@addtoreset{equation}{section}
\makeatother

\DeclareMathOperator{\Tr}{Tr}

\newcommand{\ri}{{\rm i}}

\def\th{\theta}

\def\cob{\delta}

\newcommand{\hf}{\frac{1}{2}}

\def\si{\sigma}

\def\del{\partial}

\def\lap{\Delta}
\def\bra{\langle}
\def\ket{\rangle}

\def\lrya{\leftrightarrow}

\def\la{\lambda}

\def\ka{\kappa}
\def\h#1{\widehat{#1}}
\def\bt{\beta}

\def\al{\alpha}
\def\om{\omega}

\def\rt#1{\sqrt{#1}}

\newdimen\tableauside\tableauside=1.0ex
\newdimen\tableaurule\tableaurule=0.4pt
\newdimen\tableaustep
\def\phantomhrule#1{\hbox{\vbox to0pt{\hrule height\tableaurule width#1\vss}}}
\def\phantomvrule#1{\vbox{\hbox to0pt{\vrule width\tableaurule height#1\hss}}}
\def\sqr{\vbox{%
  \phantomhrule\tableaustep
  \hbox{\phantomvrule\tableaustep\kern\tableaustep\phantomvrule\tableaustep}%
  \hbox{\vbox{\phantomhrule\tableauside}\kern-\tableaurule}}}
\def\squares#1{\hbox{\count0=#1\noindent\loop\sqr
  \advance\count0 by-1 \ifnum\count0>0\repeat}}
\def\tableau#1{\vcenter{\offinterlineskip
  \tableaustep=\tableauside\advance\tableaustep by-\tableaurule
  \kern\normallineskip\hbox
    {\kern\normallineskip\vbox
      {\gettableau#1 0 }%
     \kern\normallineskip\kern\tableaurule}%
  \kern\normallineskip\kern\tableaurule}}
\def\gettableau#1{\ifnum#1=0\let\next=\null\else
\squares{#1}\let\next=\gettableau\fi\next}

\begin{document}

\title{\bf Wilson loops in unitary matrix models at finite $N$\vskip1cm}

\author{Kazumi Okuyama}

\affiliation{Department of Physics, 
Shinshu University, Matsumoto 390-8621, Japan}

\emailAdd{kazumi@azusa.shinshu-u.ac.jp}

\abstract{
It is known that the expectation value of Wilson loops 
in the Gross-Witten-Wadia (GWW)
unitary matrix model can 
be computed exactly at finite $N$ for arbitrary representations. 
We study the perturbative and non-perturbative corrections 
of Wilson loops in the $1/N$ expansion, either 
analytically or numerically using the exact result at finite $N$.
As a by-product of the exact result of Wilson loops, 
we propose a large $N$ master field of GWW model. 
This master field has an interesting eigenvalue distribution. 
We also study the Wilson loops in large representations, called Giant Wilson loops, 
and comment on the Hagedorn/deconfinement transition 
of a unitary matrix model with a double trace interaction.
}

\maketitle

\renewcommand{\thefootnote}{\arabic{footnote}}
\setcounter{footnote}{0}
\setcounter{section}{0}

\section{Introduction\label{intro}}
Via gauge/string duality, large $N$ 't Hooft expansion of a gauge theory
corresponds to the genus expansion of dual string theory.
In general, $1/N$ expansion is an asymptotic series and
we need to include non-perturbative corrections corresponding to
various brane instantons in the bulk string theory.
We expect that we recover the exact result of gauge theory at finite $N$
after including such non-perturbative corrections. In other words, the exact result at finite $N$
can be thought of as a non-perturbative completion of the genus expansion.
We can also use this relation in the opposite direction: from the exact result at finite $N$
we can read off the information of non-perturbative corrections
either analytically or numerically.
This strategy was successfully applied to the study of instanton corrections in ABJM theory on $S^3$
from the exact values of the partition functions \cite{Hatsuda:2012hm,Hatsuda:2012dt,Hatsuda:2013gj}.
It turned out that the non-perturbative corrections in ABJM theory on $S^3$
has an interesting connection to the refined topological string on
a certain local Calabi-Yau
\cite{Hatsuda:2013oxa}.
We hope that by studying exact partition functions of gauge theories or matrix
models at finite $N$, we can reveal interesting physical/mathematical structure
of large $N$ expansion for more general cases.

In this paper, we consider the large $N$ expansion of Gross-Witten-Wadia (GWW) model \cite{Gross:1980he,Wadia:2012fr} as a simple example. The GWW model is a unitary matrix model with the
action $\Tr(U+U^\dagger)$ and it is well-known that this model has
a third order phase transition at large $N$.
Near the critical point we can take a double scaling limit \cite{Periwal:1990gf};
the GWW model in this limit describes a minimal superstring theory \cite{Klebanov:2003wg}
and the genus expansion and the non-perturbative corrections 
are well-studied in this limit.
However, somewhat surprisingly,
 the $1/N$ expansion and non-perturbative corrections
in the GWW model in the off-critical regime have not been understood completely,
and the study of such corrections from the modern viewpoint of resurgent trans-series
was initiated only recently \cite{Marino:2008ya}.
In \cite{Buividovich:2015oju,Alvarez:2016rmo}, the multi-instanton configuration
of GWW model was identified as a complex saddle of unitay matrix integral.

The GWW model is a useful testing ground to study the (non)perturbative corrections
in the large $N$ expansion since the partition function and the expectation value of Wilson loops
in arbitrary representation can be computed exactly at finite $N$.
In this paper, we study the (non)perturbative corrections in GWW model
using the exact result at finite $N$.
It is known that the genus expansion of free energy 
behaves quite differently in the two phases separated
by the third order phase transition.
In the {\it gapped} phase
where the eigenvalue density has a gap, the free energy receives all genus corrections, while
in the {\it ungapped} phase where the eigenvalue density does not have a gap,
the higher genus corrections vanish beyond genus-zero.
The ungapped phase is particularly interesting since
the instanton correction is directly accessible
by simply subtracting the genus-zero part from the exact free energy at finite $N$.  
Indeed we find a perfect agreement between the analytic 
computation of instanton correction and the exact free energy at finite $N$. 

We can study the expectation value of winding Wilson loops $\bra \Tr U^k\ket$
with winding number $k=1,2,\cdots$, in a similar manner.
In the gapped phase we compare the exact result and the genus
expansion of matrix model and find a perfect agreement. 
In the ungapped phase, $\bra \Tr U^k\ket$ with $k\geq2$ has no perturbative correction
and hence the instanton correction is directly accessible.
We determine the 
coefficient of instanton correction from numerical fitting using the exact result at finite $N$.

We also consider the so-called Giant Wilson loops in the large (anti)symmetric representation,
where the rank of the representation becomes of order $N$ \cite{Grignani:2009ua,Karczmarek:2010ec,Karczmarek:2011gk}.
We compute the one-loop correction to the leading large $N$
result of Giant Wilson loops obtained in \cite{Grignani:2009ua,Karczmarek:2010ec,Karczmarek:2011gk},
and we find that the matching with the exact result is improved 
by adding the one-loop correction.

As an interesting by-product of exact result of Wilson loops,
we propose a ``master field'' of GWW model.
The exact form of $\bra \Tr U\ket$
in \eqref{eq:exact-tr}
and $\bra \det(x-U)\ket$ in \eqref{eq:exact-ch}
suggests that the $N\times N$ matrix $M_0^{-1}M_1$,
with $M_k$ defined in \eqref{eq:Mk},
can be thought of as a master field of GWW model.
It turns out that this master field has an interesting distribution of eigenvalues.
In particular, we find numerically that in the ungapped phase
the eigenvalues of master field are distributed along a contour of constant effective potential,
and this contour is located inside the unit circle on a complex
plane.

As another example, we study the free energy and (Giant) Wilson loops in a unitary matrix model
with a double-trace interaction $\Tr U\Tr U^\dagger$, which we call the ``adjoint model''.
This model 
naturally appears as a truncation of the thermal partition function
of $d=4$ $\mathcal{N}=4$ super Yang-Mills (SYM) theory on $S^3\times S^1$ \cite{Sundborg:1999ue}.  
This model exhibits a Hagedorn/deconfinement transition, which is 
holographically dual to the Hawking-Page transition on the bulk gravity side \cite{Witten:1998zw}.
As discussed in \cite{Liu:2004vy},
we can compute the partition function and Wilson loops in the adjoint model by
a certain integral transformation of the GWW model.
Using this relation to the GWW model, we study numerically the behavior of
partition function and Wilson loops in the adjoint model.

This paper is organized as follows. 
In section \ref{sec:free} we study the free energy of GWW model.
We find that the exact partition function at finite $N$
correctly reproduces the analytic results of the large $N$ expansion
of free energy in both gapped phase and the ungapped phase.  
In section \ref{sec:wind} we study the winding Wilson loops $\bra \Tr U^k\ket$ 
in GWW model. In the gapped phase we find that the exact result at finite $N$
reproduces the analytic result of genus expansion.
In the ungapped phase we determine the coefficients of the first non-trivial instanton 
correction by numerical fitting.
In section \ref{sec:master} we propose a master field of GWW model and study its
eigenvalue distribution. In the gapped phase eigenvalues of the master field approaches
the known distribution in \cite{Gross:1980he,Wadia:2012fr} 
as $N$ becomes large, while in the ungapped phase
we find that the eigenvalues of the master field are distributed inside the unit circle.
In section \ref{sec:conn} using the exact form of the Wilson loops in general representations,
we study the connected part of multi-trace expectation values.
In section \ref{sec:giant} we study the Wilson loops in the $k$-th (anti)symmetric representation
in the limit where $k,N\to\infty$ with $k/N$ fixed.
In section \ref{sec:adj} we study the adjoint model with a double-trace
interaction $\Tr U\Tr U^\dagger$.
We consider the free energy, winding Wilson loops, and Giant Wilson loops in the
adjoint model, and study the behavior of these quantities 
under the Hagedorn/deconfinement transition.
We conclude in section \ref{sec:discuss}  with some discussions and future directions.
In addition, we have four appendices.
In appendix \ref{app:exact}, we review the exact result
of the partition function and Wilson loops in GWW model at finite $N$.
In appendix \ref{app:potential}, we compute the 
effective potential for a probe eigenvalue in the ungapped phase of GWW model.
In appendix \ref{app:bessel}, we study the one-instanton correction 
in the ungapped phase of GWW model and determine the overall coefficient of instanton correction 
by matching the result of double-scaling limit.
In appendix \ref{app:resolvent}, we compute the genus-one resolvent 
of GWW model in the gapped phase by using the mapping between the unitary matrix model and 
the hermitian matrix model.

\section{Free energy of GWW model \label{sec:free}}
We are interested in the non-perturbative corrections
in the large $N$
expansion of the GWW model defined by\footnote{Note that our convention of coupling constant is different from \cite{Marino:2008ya}
\begin{align}
 Z=\int_{U(N)} dU \exp\left[\frac{1}{2g_s}\Tr(U+U^\dagger)\right],
\label{eq:Z-gs}
\end{align}
where the string coupling $g_s$ and the 't Hooft coupling $t=Ng_s$ are related to 
our coupling $g$ by
\begin{align}
  g_s=\frac{1}{Ng},\qquad g=\frac{1}{t}.
\label{eq:gs-t}
\end{align}
\label{foot:Z-gs}
}
\begin{align}
 Z(N,g)=\int_{U(N)} dU \exp\left[\frac{Ng}{2}\Tr(U+U^\dagger)\right].
\end{align}
It is well-known that the partition function
of GWW model can be evaluated exactly at finite $N$ \cite{Wadia:2012fr,Bars:1979xb} 
\footnote{See  \cite{Rossi:1996hs} for a review of unitary matrix models.}
\begin{align}
Z(N,g)= \det\Bigl[I_{i-j}(Ng)\Bigr]_{i,j=1,\cdots N},
\label{eq:Z-exact}
\end{align}
where $I_\nu(x)$ denotes the modified Bessel function of the first kind.
As we will see below, we can study perturbative and non-perturbative corrections to the free
energy in the large $N$ expansion from the exact
result at finite $N$ \eqref{eq:Z-exact}.

In the large $N$ limit with fixed $g$,
the free energy admits the genus expansion
\begin{align}
 \log Z(N,g)=\sum_{\ell=0}^\infty N^{2-2\ell}F_\ell(g)
+F^{(\text{inst})}
\label{eq:free-genus-expand}
\end{align}
where $F^{(\text{inst})}$ denotes the exponentially suppressed correction
\begin{align}
 F^{(\text{inst})}=\mathcal{O}(e^{-N}).
\end{align}
As shown in the seminal papers \cite{Gross:1980he,Wadia:2012fr}
there is a third order phase transition at  $g=1$
and the
genus-zero free energy  behaves differently below and above the transition point
$g=1$
\begin{align}
F_0(g)=\left\{
\begin{aligned}
 &\frac{g^2}{4},& \quad (g<1),\\
&g-\hf\log g-\frac{3}{4},& \quad (g>1).
\end{aligned}
\right.
\label{eq:F0-GW}
\end{align}
This third order phase transition is associated with the opening/closing of the gap
of the distribution of eigenvalue $e^{\ri\th}$
of unitary matrix $U$.
The eigenvalue density $\rho(\th)$ has no gap when $g<1$ ({\em ungapped phase})
while it has a gap when $g>1$ ({\em gapped phase}):
\begin{align}
\rho(\th)=\left\{
\begin{aligned}
 &\frac{1}{2\pi}(1+g\cos\th),& \quad (|\th|\leq\pi, ~~g<1), \\
&\frac{g}{\pi}\cos\frac{\th}{2}\rt{\frac{1}{g}-\sin^2\frac{\th}{2}},& 
\quad (|\th|\leq\al, ~~g>1).
\end{aligned}
\right.
\label{eq:rho}
\end{align} 
Here $\al$ is the end-point of eigenvalue distribution given by
\begin{align}
 \al=2\arcsin(g^{-1/2}).
\end{align}

\begin{figure}[tbh]
\centering\includegraphics[width=8cm]{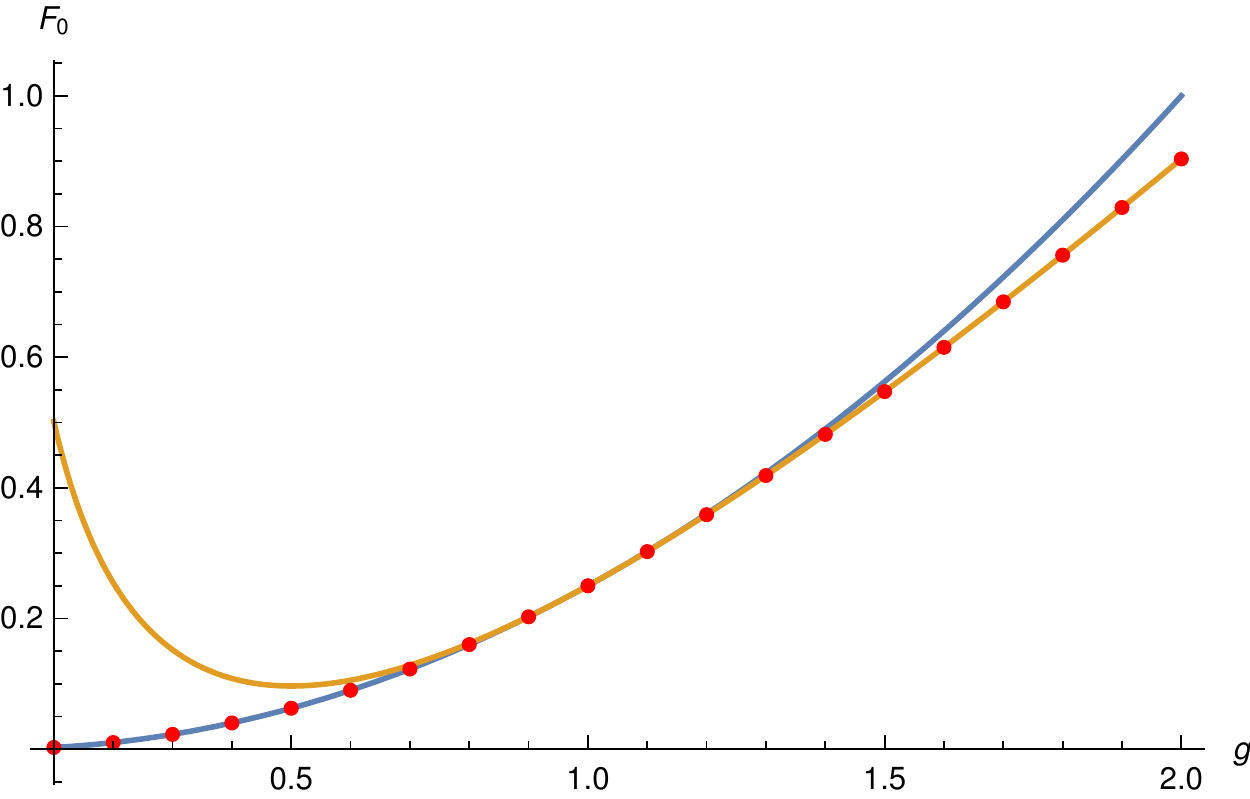}
  \caption{
Plot of the genus-zero free energy $F_0(g)$. 
The red dots are the exact value of the free energy $\frac{1}{N^2}\log Z(N,g)$
for $N=100$,
while the blue curve and the orange curve represent the analytic form of
$F_0(g)$ in \eqref{eq:F0-GW}
in the ungapped phase  and the gapped phase, respectively.
}
  \label{fig:F0}
\end{figure}
In Fig.~\ref{fig:F0}, we plot the genus-zero free energy in \eqref{eq:F0-GW}
and the exact free energy for $N=100$ and find a nice agreement, as expected.

\subsection*{Perturbative corrections in the gapped phase}

In the gapped phase $(g>1)$, we can systematically compute 
the genus-$\ell$ free energy
by solving the so-called {\it pre-string}  equation
obtained from the method of orthogonal polynomials \cite{Goldschmidt:1979hq,Periwal:1990gf,Marino:2008ya}.
The first three terms are given by
\begin{align}
 \begin{aligned}
  F_1(g)&=\zeta'(-1)-\frac{1}{12}\log N
-\frac{1}{8}\log(1-1/g),\\
F_2(g)&=-\frac{1}{240}+\frac{3}{128(g-1)^3},\\
F_3(g)&=\frac{1}{1008}+\frac{9(5g+2)}{1024(g-1)^6}.
 \end{aligned}
\label{eq:Fell-anal}
\end{align}
In general, the genus $\ell$
free energy $F_\ell(g)$ has a structure
\begin{align}
 F_\ell(g)=\frac{B_{2\ell}}{2\ell(2\ell-2)}+\frac{1}{(g-1)^{3\ell-3}}\sum_{n=0}^{\ell-2}c_n^{(\ell)}g^n
\label{eq:Fg-general}
\end{align}
where $B_{2\ell}$ 
denotes the Bernoulli number which comes from the volume of $U(N)$ gauge group.

\begin{figure}[htb]
\centering
\subcaptionbox{$F_1(g)$\label{fig:F1}}{\includegraphics[width=4.5cm]{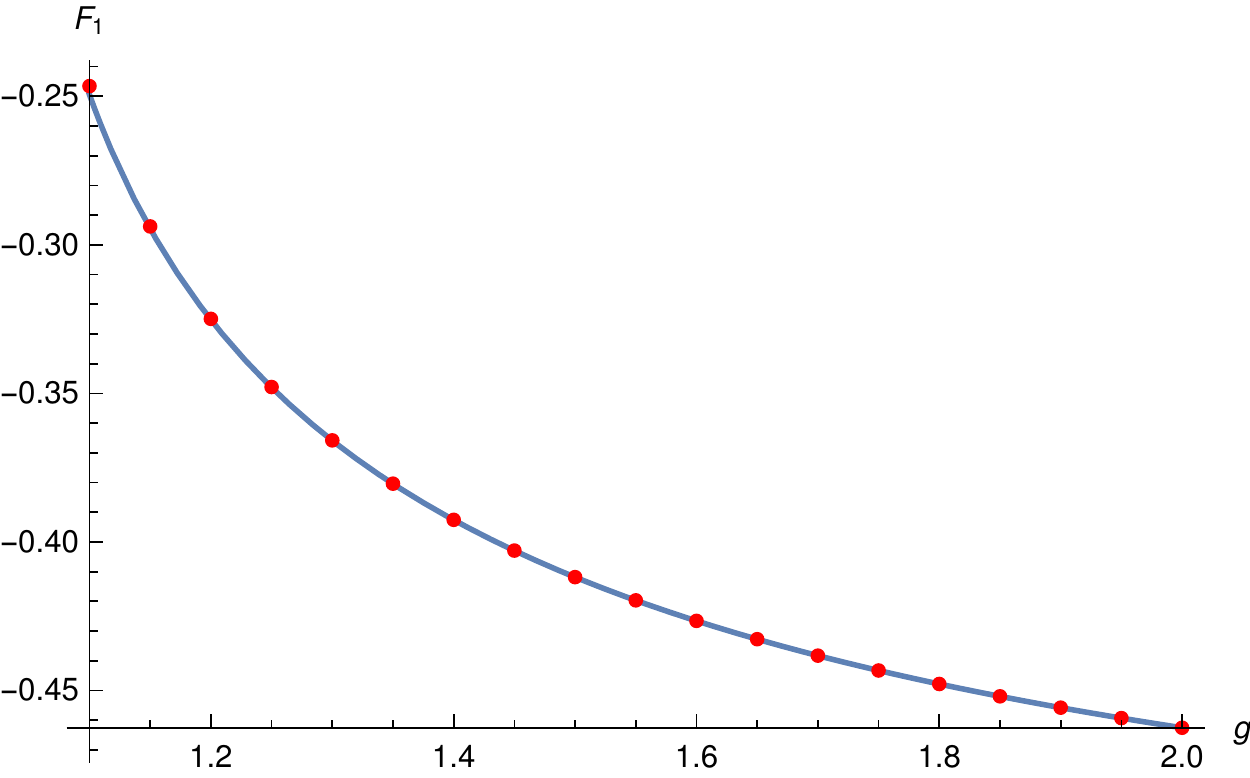}}
\hskip2mm
\subcaptionbox{$F_2(g)$\label{fig:F2}}{\includegraphics[width=4.5cm]{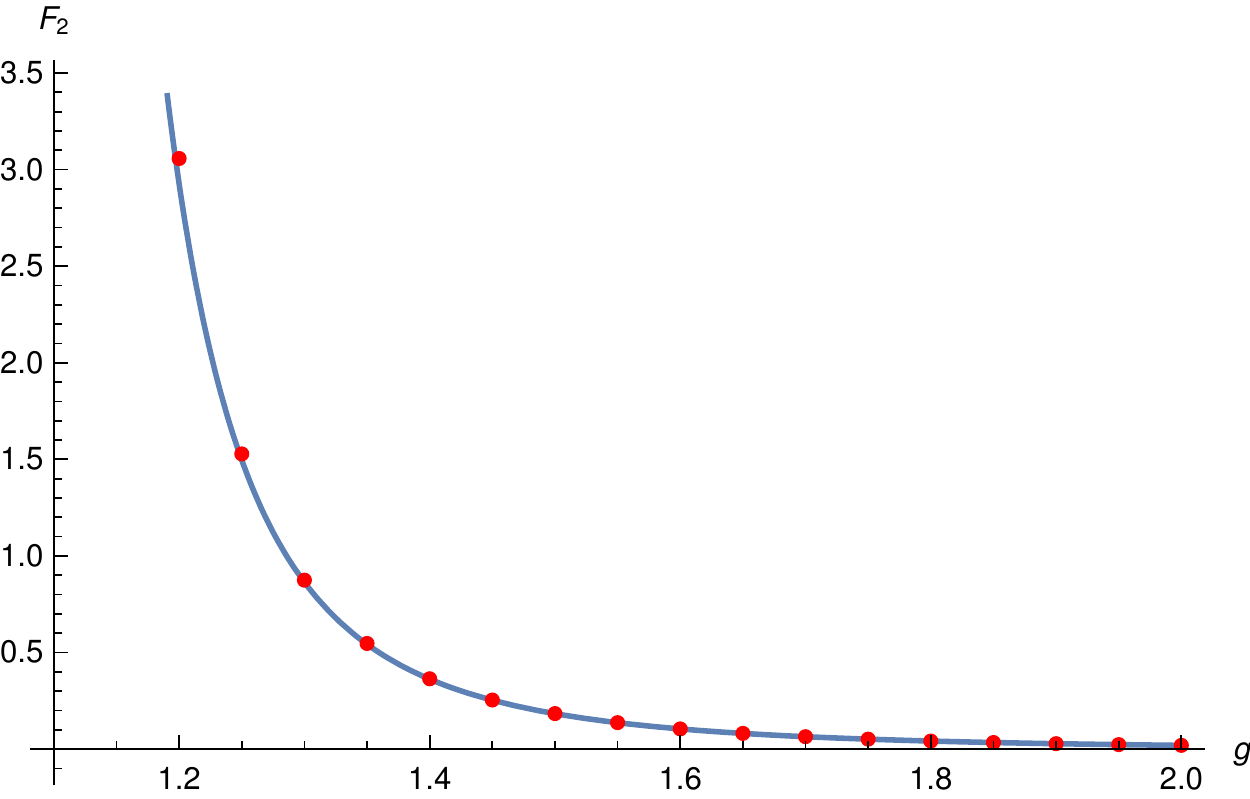}}
\hskip2mm
\subcaptionbox{$F_3(g)$\label{fig:F3}}{\includegraphics[width=4.5cm]{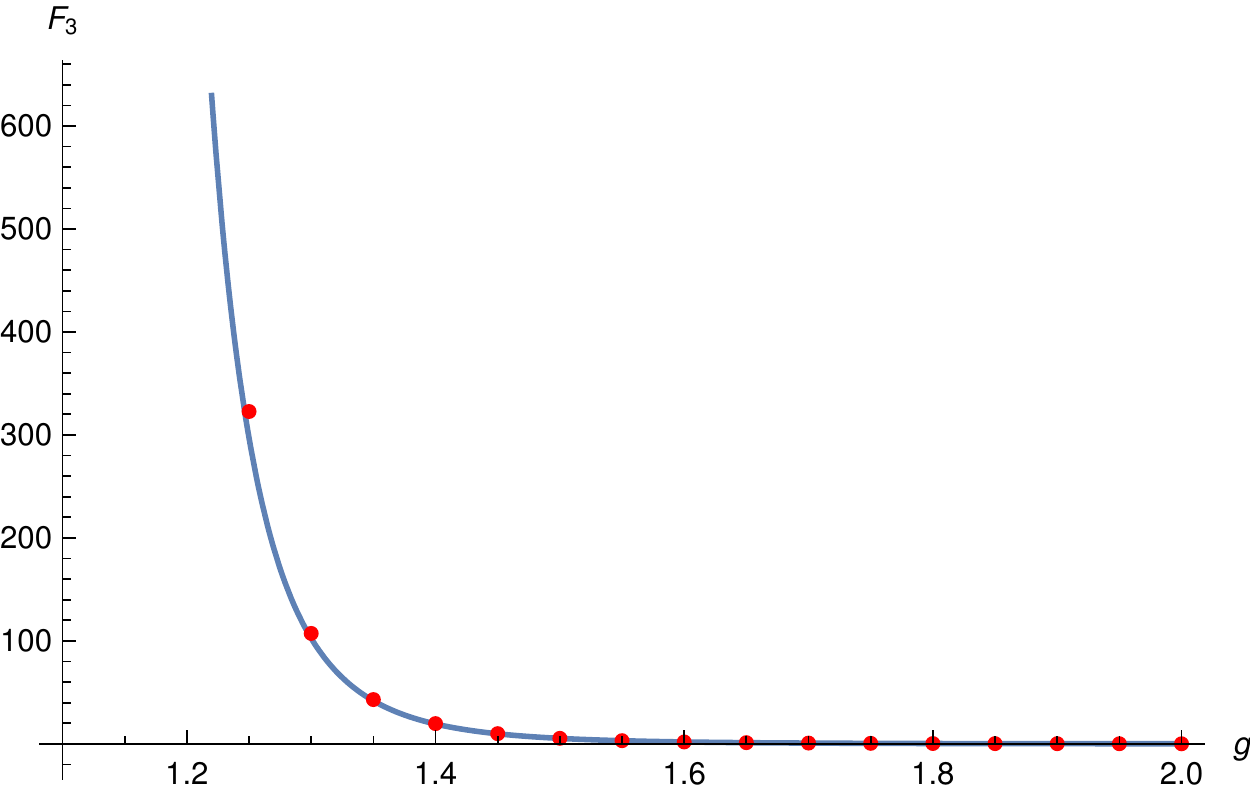}}
  \caption{
Plot 
of the genus-$\ell$ free energy $F_\ell(g)$ for $\ell=1,2,3$
in the gapped phase $(g>1)$.
The dots are the values obtained from the exact free energy $\log Z(N,g)$
for $N=100$
using \eqref{eq:exact-Fell}, while
solid curves represent the analytic form of $F_\ell(g)$ in \eqref{eq:Fell-anal}.
}
  \label{fig:F123}
\end{figure}

One can extract the genus-$\ell$ free energy from the exact value of $Z(N,g)$
in \eqref{eq:Z-exact}
by subtracting the lower genus contributions
\begin{align}
 F_\ell(g)\approx N^{2\ell-2}\left(\log Z(N,g)-\sum_{\ell'=0}^{\ell-1}N^{2-2\ell'}F_{\ell'}(g)\right),
\quad(N\gg1).
\label{eq:exact-Fell}
\end{align}
As we can see from Fig.~\ref{fig:F123}, the exact partition function \eqref{eq:Z-exact}
nicely matches the analytic result of genus-$\ell$ free energy \eqref{eq:Fell-anal}
as expected.

The instanton correction in the gapped phase has been studied in \cite{Marino:2008ya}.
The genus expansion in the gapped phase is Borel non-summable and 
in order to compare with the exact result at finite $N$ we need to add the lateral
Borel resummations along the integration contours below and above the real axis.
On the other hand, in the ungapped phase the perturbative
genus expansion stops at first order and 
we do not need to perform the
Borel resummation of perturbative part.
As a consequence, in the ungapped phase we can directly access to the 
instanton correction from the exact result at finite $N$, as we will see below.

\subsection*{Instanton correction in the ungapped phase}

In the ungapped phase ($g<1$),
the genus expansion of free energy stops at genus-zero
\begin{align}
 F_0(g)=\frac{g^2}{4},\qquad F_\ell(g)=0~~~(\ell\geq1),
\end{align}
and the instanton correction starts from the two-instanton correction 
\footnote{As explained in appendix \ref{app:bessel},
the expectation value of $\det U$ receives one-instanton correction $\mathcal{O}(e^{-NS_\text{inst}(g)})$, while the instanton correction to the free energy starts from the two-instanton 
$\mathcal{O}(e^{-2NS_\text{inst}(g)})$.}
\begin{align}
 F^{(\text{inst})}=e^{-2NS_\text{inst}(g)}\sum_{n=1}^\infty \frac{f_n(g)}{N^n}
+\mathcal{O}(e^{-4NS_\text{inst}(g)}),
\label{eq:inst-co}
\end{align}
where the instanton action is given by \cite{Green:1981mx,Rossi:1982vw}
\begin{align}
 S_\text{inst}(g)=\cosh^{-1}(1/g)-\rt{1-g^2}.
\label{eq:Sinst-g}
\end{align}

\begin{figure}[tbh]
\centering\includegraphics[width=8cm]{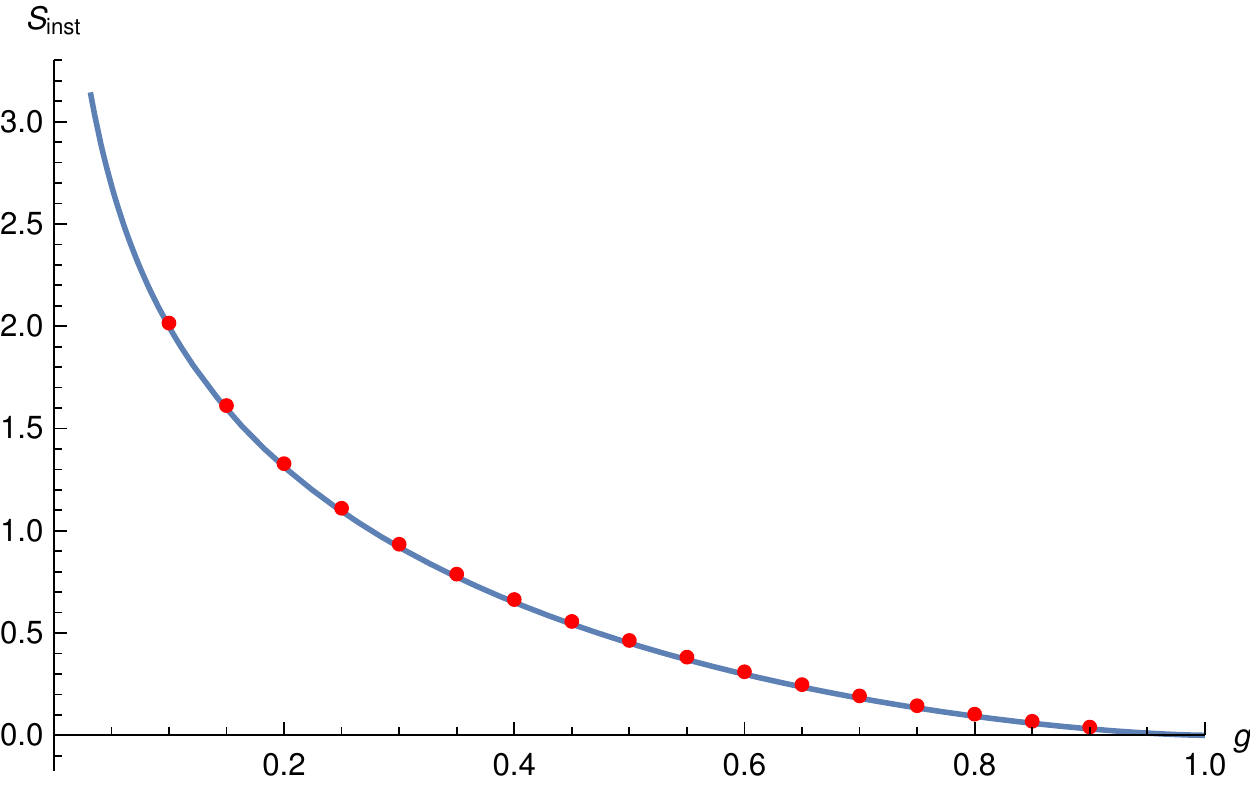}
  \caption{
Plot of the instanton action $S_\text{inst}(g)$ in the range $0<g<1$.
The red dots are the numerical values extracted from the 
exact free energy using \eqref{eq:Sinst-approx} with $N=400$,
while the solid curve represent the analytic form of $S_\text{inst}(g)$ in \eqref{eq:Sinst-g}.
}
  \label{fig:Sinst}
\end{figure}

One can extract the instanton action numerically from the exact partition function
$Z(N,g)$ by subtracting the perturbative part
\begin{align}
S_\text{inst}(g)\approx  -\frac{1}{2N}\log\Biggl|\log Z(N,g)-N^2F_0(g)\Biggr|
,\qquad (N\gg1)
\label{eq:Sinst-approx}
\end{align}
As shown in Fig.~\ref{fig:Sinst}, the exact $Z(N,g)$
correctly reproduces the analytic result of instanton action \eqref{eq:Sinst-g}.

As explained in appendix \ref{app:bessel}, 
we can systematically 
compute the instanton coefficient $f_n$ in \eqref{eq:inst-co}
\begin{align}
\begin{aligned}
 F^{(\text{2-inst})}
=\frac{e^{-2NS_\text{inst}(g)}}{8\pi N}\left[
-\frac{g^2}{(1-g^2)^{\frac{3}{2}}}+\frac{1}{N}\frac{g^2(26+9g^2)}{12(1-g^2)^3}
-\frac{1}{N^2}\frac{g^2(297g^4+2484g^2+964)}{288(1-g^2)^{\frac{9}{2}}}+\cdots\right] 
\end{aligned}
\label{eq:inst-ungap}
\end{align}
Instanton coefficient in the ungapped phase has been studied in \cite{Marino:2008ya}
but the overall factor was not determined in \cite{Marino:2008ya}.
We have fixed 
the overall factor ($1/8\pi N$ in \eqref{eq:inst-ungap})
by matching the Hastings-McLeod solution of Painlev\'{e} II equation 
in the double scaling limit (see appendix \ref{app:bessel} for details). 
Also, we have checked numerically that the 
instanton correction \eqref{eq:inst-ungap} to the free energy
correctly reproduces the exact value of $\log Z(N,g)-N^2F_0(g)$.

\section{Winding Wilson loops \label{sec:wind}}
In this section, we consider the expectation value of winding
Wilson loop $\bra \Tr U^k\ket$
with winding number $k\in \mathbb{Z}_{>0}$.
One can show that $\bra \Tr U^k\ket$ can be computed exactly at finite $N$
(see appendix \ref{app:exact} for a derivation)
\begin{align}
\bra \Tr U^k \ket=
\Tr (M_0^{-1}M_k),
\label{eq:exact-tr}
\end{align} 
where $M_k$ is an $N\times N$ matrix
whose $(i,j)$ element is given by
\begin{align}
 (M_k)_{i,j}=I_{k+i-j}(Ng),\qquad (i,j=1,\cdots,N).
\label{eq:Mk}
\end{align}
For $k=1$ the expectation value is related to the derivative of free energy
\begin{align}
 \frac{1}{N}\bra \Tr U\ket=\frac{1}{N^2}\del_g \log Z(N,g).
\label{eq:u1-del}
\end{align}
In the planar limit we find
\begin{align}
 \frac{1}{N}\bra \Tr U\ket=\del_g F_0(g)=
\left\{
\begin{aligned}
 &\frac{g}{2},\qquad &(g<1),\\
&1-\frac{1}{2g},\qquad &(g>1).
\end{aligned}\right.
\label{eq:u1-vev}
\end{align}
For $k\geq2$ the expectation value in the planar limit is obtained using the 
eigenvalue density \eqref{eq:rho} as
\begin{align}
 \frac{1}{N}\bra \Tr U^k\ket=\int d\th \rho(\th)e^{\ri k\th}=
\left\{
\begin{aligned}
 &0,\qquad &(g<1),\\
&\frac{1}{k-1}\Bigl(1-\frac{1}{g}\Bigr)^2 P^{(1,2)}_{k-2}\Bigl(1-\frac{2}{g}\Bigr),\qquad &(g>1),
\end{aligned}\right.
\label{eq:uk-lead}
\end{align}
where $P^{(\al,\bt)}_n(x)$ denotes the Jacobi polynomial.
\begin{figure}[tbh]
\centering\includegraphics[width=8cm]{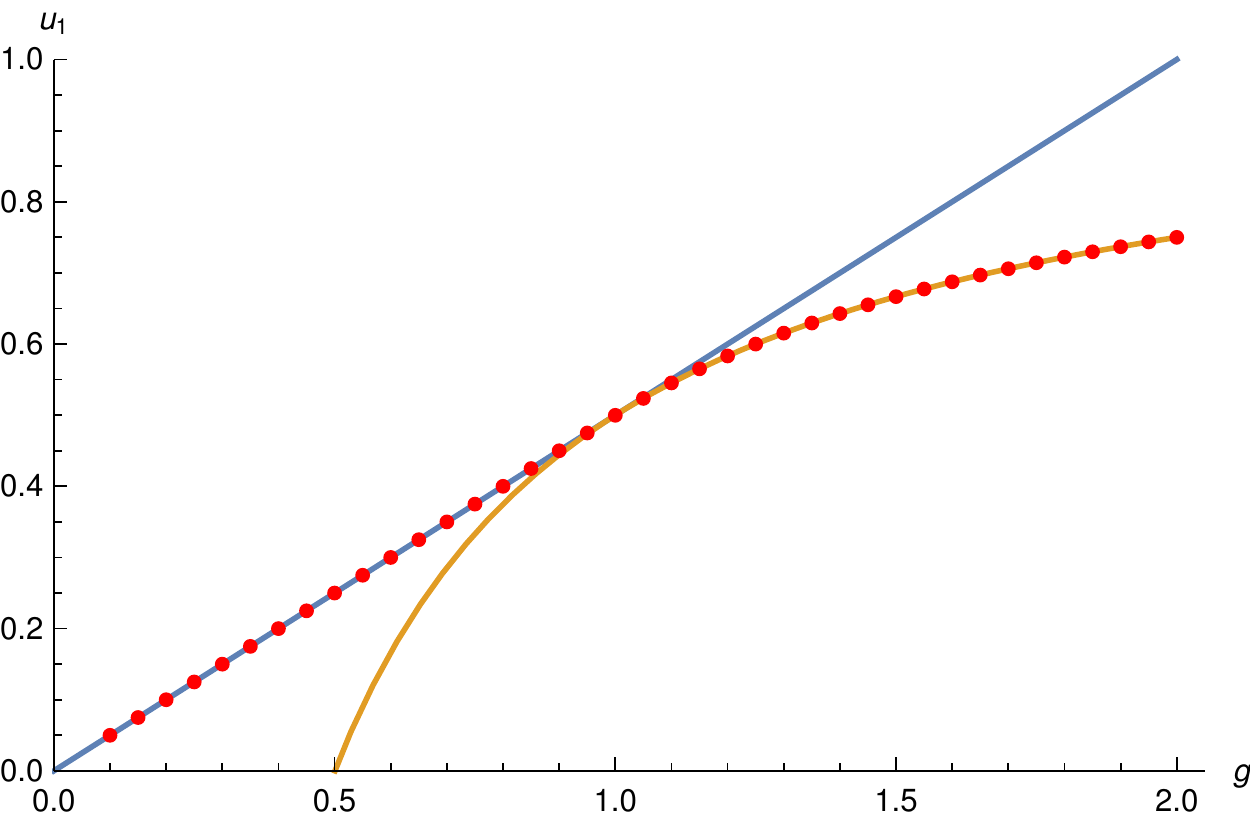}
  \caption{
Plot of the expectation value of Wilson loop $u_1=\frac{1}{N}\bra \Tr U\ket$.
The red dots are the exact value for $N=100$, while
the blue curve and the orange curve are
the planar limit \eqref{eq:u1-vev} in the ungapped phase and the gapped phase, respectively.
}
  \label{fig:w1}
\end{figure}

\begin{figure}[htb]
\centering
\subcaptionbox{$u_2$\label{fig:F1}}{\includegraphics[width=4.5cm]{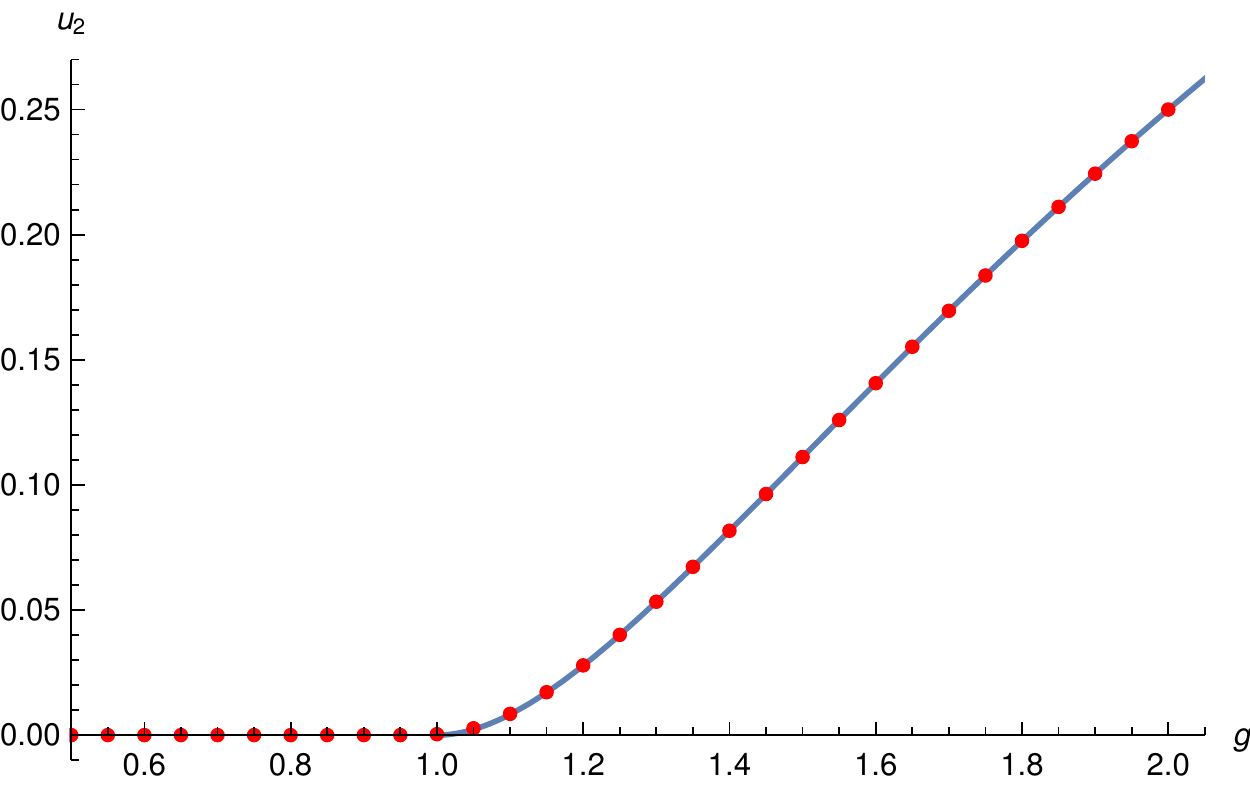}}
\hskip8mm
\subcaptionbox{$u_3$\label{fig:F2}}{\includegraphics[width=4.5cm]{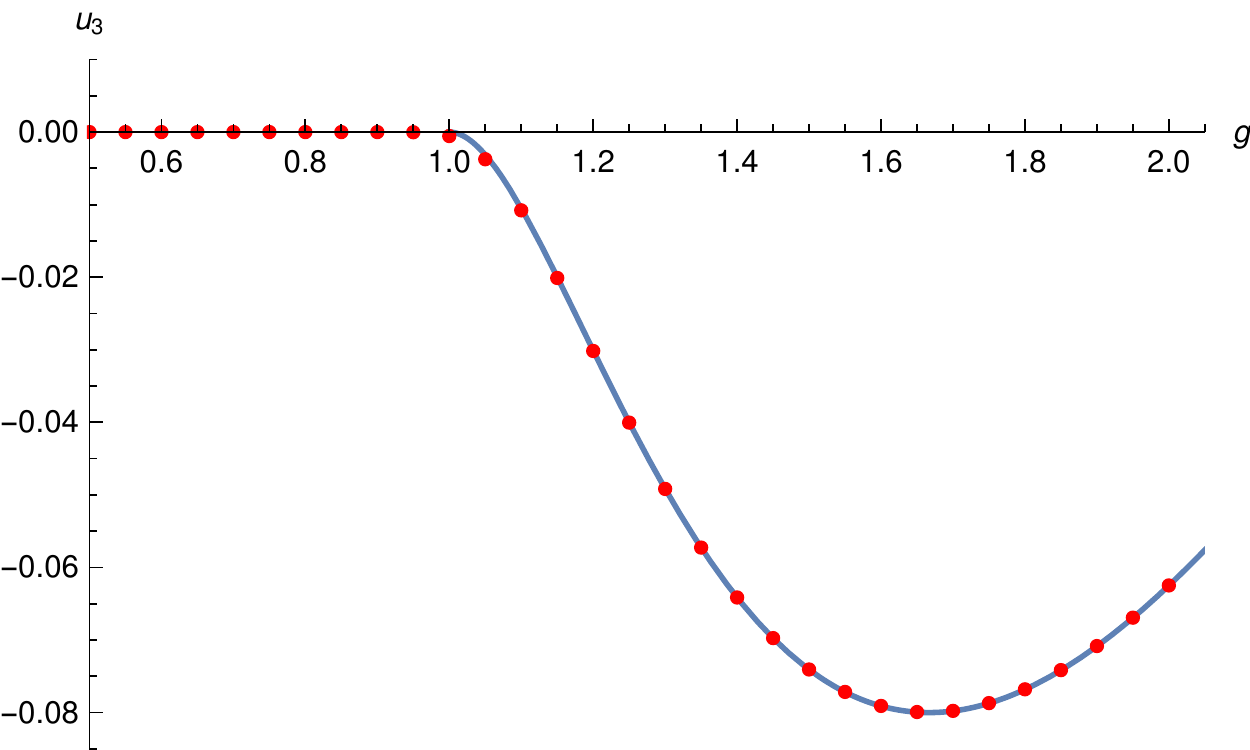}}
\hskip8mm
\subcaptionbox{$u_4$\label{fig:F3}}{\includegraphics[width=4.5cm]{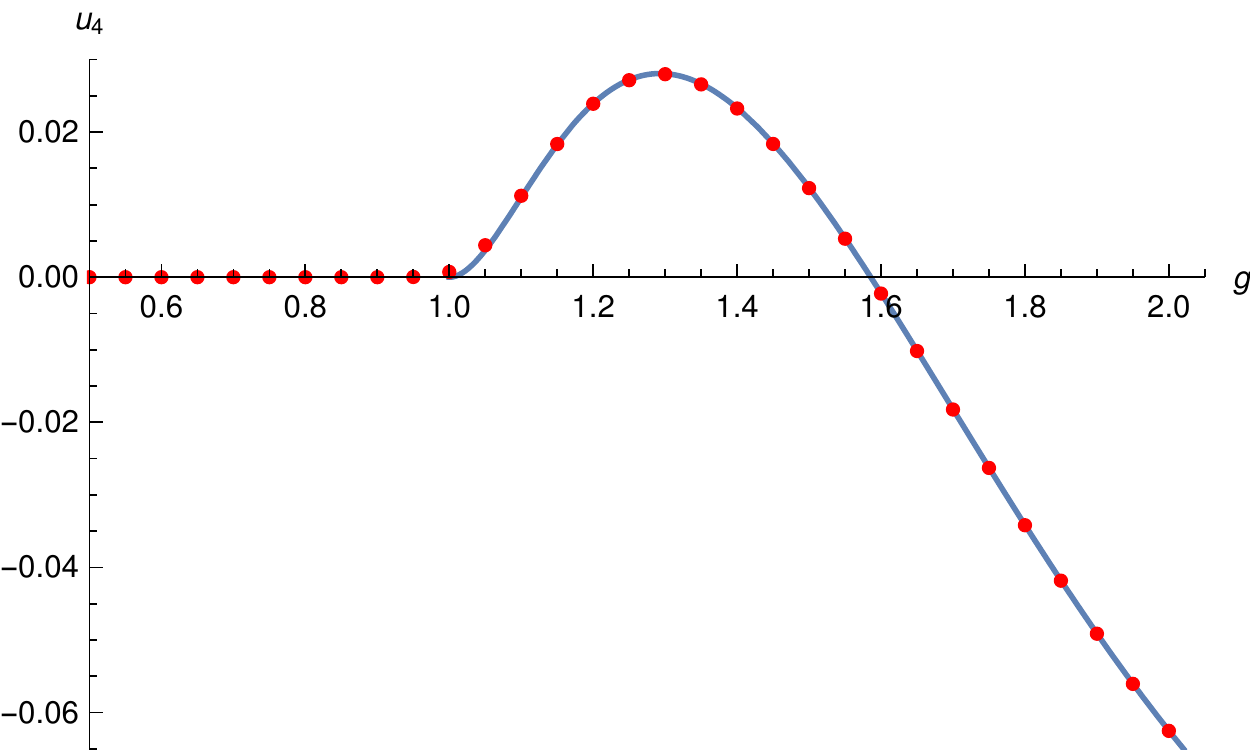}}
  \caption{
Plot 
of the expectation value of the winding Wilson loops 
$u_k=\frac{1}{N}\bra \Tr U^k\ket$ for $k=2,3,4$.
The red dots are the exact values
at $N=100$, while
solid curves represent the planar limit in 
\eqref{eq:uk-lead}.
}
  \label{fig:w234}
\end{figure}
Again, we can compare the analytic expression of $\frac{1}{N}\bra \Tr U^k\ket$ in
the planar limit \eqref{eq:uk-lead} and the exact value at finite $N$ \eqref{eq:exact-tr}.
In Fig.~\ref{fig:w1} and   Fig.~\ref{fig:w234}, we show the 
plot of $\frac{1}{N}\bra \Tr U^k\ket$ for $k=1,2,3,4$. We find perfect agreement between the
analytic result and the exact value at finite $N$, as expected.

\subsection*{Genus expansion in the gapped phase}
From the exact value of winding Wilson loops at finite $N$ \eqref{eq:exact-tr}, 
one can determine the higher genus correction to the winding Wilson loops
by numerical fitting. In the gapped phase,
winding Wilson loops receives all-order corrections in the $1/N$ expansion.
For winding numbers $k=1,\cdots,5$, 
we find numerically the genus expansion in the gapped phase:
\begin{align}
 \begin{aligned}
  \frac{1}{N}\bra \Tr U\ket&=1-\frac{1}{2g}
-\frac{1}{N^2}\frac{1}{8(g-1)g}-\frac{1}{N^4}\frac{9}{128(g-1)^4}
-\frac{1}{N^6}\frac{9 (25 g+17)}{1024 (g-1)^7}+\cdots,\\
\frac{1}{N}\bra \Tr U^2\ket&=\frac{(g-1)^2}{g^2}
+\frac{1}{N^2}\frac{1}{4(g-1)g^2}+\frac{1}{N^4}\frac{9}{64(g-1)^4g}
+\frac{1}{N^6}\frac{451 g^2+297 g+23}{1024 (g-1)^7 g^2}+\cdots,\\
 \frac{1}{N}\bra \Tr U^3\ket&=\frac{(g-1)^2(2g-5)}{2g^3}
+\frac{1}{N^2}\frac{10-28g+15g^2}{8(g-1)g^3}
+\frac{1}{N^4}\frac{3(20-90g+96g^2-35g^3)}{128(g-1)^4g^3}+\cdots,\\
 \frac{1}{N}\bra \Tr U^4\ket&=\frac{(g-1)^2 \left(g^2-6 g+7\right)}{g^4}
+\frac{1}{N^2}\frac{-35+90 g -70 g^2+16 g^3}{2 (g-1) g^4}\\
&+\frac{1}{N^4}\frac{-154+561 g-624 g^2+226 g^3}{32 (g-1)^4 g^4}+\cdots,\\
\frac{1}{N}\bra \Tr U^5\ket&=\frac{(g-1)^2 \left(2 g^3-21 g^2+56 g-42\right)}{2 g^5}
+\frac{1}{N^2}\frac{5 \left(35 g^4-260 g^3+630 g^2-616 g+210\right)}{8
   (g-1) g^5}\\
&+\frac{1}{N^4}\frac{26460 g^5-130688 g^4+241751 g^3-209326 g^2+84392
   g-12772}{512 (g-1)^4 g^5}+\cdots.
 \end{aligned}
\label{eq:trU-genus}
\end{align}
The planar part of \eqref{eq:trU-genus}
agrees with \eqref{eq:u1-vev} for $k=1$ and \eqref{eq:uk-lead}
for $k\ge2$.
One can in principle compute the higher genus corrections of winding Wilson loops
analytically and compare our numerical result \eqref{eq:trU-genus}.
For instance, the
genus-one resolvent can be easily found by mapping the unitary matrix model to
a hermitian matrix model by a change of variable \cite{Mizoguchi:2004ne}.
As explained in appendix \ref{app:resolvent},
we have checked that the genus-one correction in \eqref{eq:trU-genus}
is correctly reproduced from the analytic form of the
genus-one resolvent.
It would be interesting to analytically compute the  higher genus corrections to
the winding Wilson loops
and compare
our numerical result \eqref{eq:trU-genus}. 

\subsection*{Instanton correction in the ungapped phase}
Let us consider the instanton correction
to the winding Wilson loop
$\bra \Tr U^k\ket$ in the ungapped phase $(g<1)$.
For $k=1$, the two-instanton correction is readily obtained by 
taking the derivative of  free energy with respect to $g$ \eqref{eq:u1-del}
\begin{align}
  \frac{1}{N}\bra \Tr U\ket -\frac{g}{2}=\frac{e^{-2NS_\text{inst}(g)}}{4\pi N^2}
\left[\frac{-g}{1-g^2}+\frac{1}{N}\frac{g(14+3g^2)}{12(1-g^2)^{\frac{5}{2}}}
-\frac{1}{N^2}\frac{g(340+804g^2+81g^4)}{288(1-g^2)^{4}}
+\mathcal{O}(N^{-3})\right] .
\end{align}
For $k\geq2$, there is no perturbative piece and the non-zero contribution starts from
the two-instanton correction. From the exact value of
$\bra \Tr U^k\ket$ in \eqref{eq:exact-tr}, we determined
the instanton coefficients by numerical fitting
\begin{align}
 \begin{aligned}
 \frac{1}{N}\bra \Tr U^2\ket &= \frac{e^{-2NS_\text{inst}(g)}}{4\pi N^2}
\left[\frac{2}{1-g^2}-\frac{1}{N}\frac{28+5g^2}{12(1-g^2)^{\frac{5}{2}}}+\mathcal{O}(N^{-2})\right],\\
\frac{1}{N}\bra \Tr U^3\ket &=\frac{e^{-2NS_\text{inst}(g)}}{4\pi N^2} 
\left[\frac{-4+g^2}{(1-g^2)g}+\mathcal{O}(N^{-1})\right],\\
\frac{1}{N}\bra \Tr U^4\ket &= \frac{e^{-2NS_\text{inst}(g)}}{4\pi N^2}
\left[\frac{8-4g^2}{(1-g^2)g^2}+\mathcal{O}(N^{-1})\right].
 \end{aligned}
\label{eq:trU-inst}
\end{align}
As far as we know, no systematic method to compute instanton corrections for 
general Wilson loops is known in the literature.
It would be interesting to develop a technique to compute instanton corrections
to the Wilson loops and see if our numerical results \eqref{eq:trU-inst} are reproduced.

\section{Master field of GWW model and its eigenvalue distribution \label{sec:master}}
In this section we propose a ``master field'' of GWW model and study its eigenvalue
distribution.
\subsection*{Master field of GWW model}
From the relation $\bra\Tr U\ket=\Tr (M_0^{-1}M_1)$ in \eqref{eq:exact-tr},
it is natural to conjecture that the $N\times N$ matrix $M_0^{-1}M_1$
can be thought of as a ``master field'' of GWW model
\begin{align}
 U~\lrya~M_0^{-1}M_1.
\end{align}
In fact, we can prove more general correspondence:
expectation value of the characteristic polynomial
of $U$ is given by the characteristic polynomial
of master field (see appendix \ref{app:exact})
\begin{align}
 \bra\det(x-U)\ket=\det(x-M_0^{-1}M_1).
\label{eq:exact-ch}
\end{align}
Moreover, we have checked numerically that
the expectation values of winding Wilson loops are also reproduced from
the trace of master field in the large $N$ limit
\begin{align}
 \bra \Tr U^k\ket=\Tr(M_0^{-1}M_1)^k,\qquad (N\gg1).
\label{eq:master-wind}
\end{align}
Note that for $k=1$ the relation \eqref{eq:master-wind} is exact  at finite $N$,
while for $k\geq2$ this relation \eqref{eq:master-wind} holds only in the planar limit.

From the explicit form of the matrix $M_k$ \eqref{eq:Mk},
one can easily show that the master field $M_0^{-1}M_1$ 
has the form
\footnote{We would like to thank Pavel Buividovich for pointing out this structure.}
\begin{align}
 M_0^{-1}M_1=
\begin{pmatrix}
 a_1 & 1 & 0  & \ldots & 0 \\ 
a_2 & 0 & 1  & \ldots & 0 \\
\vdots & & & \ddots   & \vdots\\
a_{N-1} & 0 & 0  & \ldots  & 1 \\
a_N & 0 & 0 &  \ldots & 0
\end{pmatrix},
\end{align}
where $a_i$ appears as the coefficient of characteristic polynomial
\begin{align}
 \bra \det(x-U)\ket=x^N -\sum_{i=1}^N a_i x^{N-i}.
\end{align}
In other words, $a_i$ is the expectation value of 
Wilson loops in the $i$-th anti-symmetric representation
up to a sign $(-1)^{i-1}$.

\subsection*{Eigenvalue distribution of master field}
\begin{figure}[htb]
\centering
\subcaptionbox{$g=1.1$\label{fig:g11}}{\includegraphics[width=4.5cm]{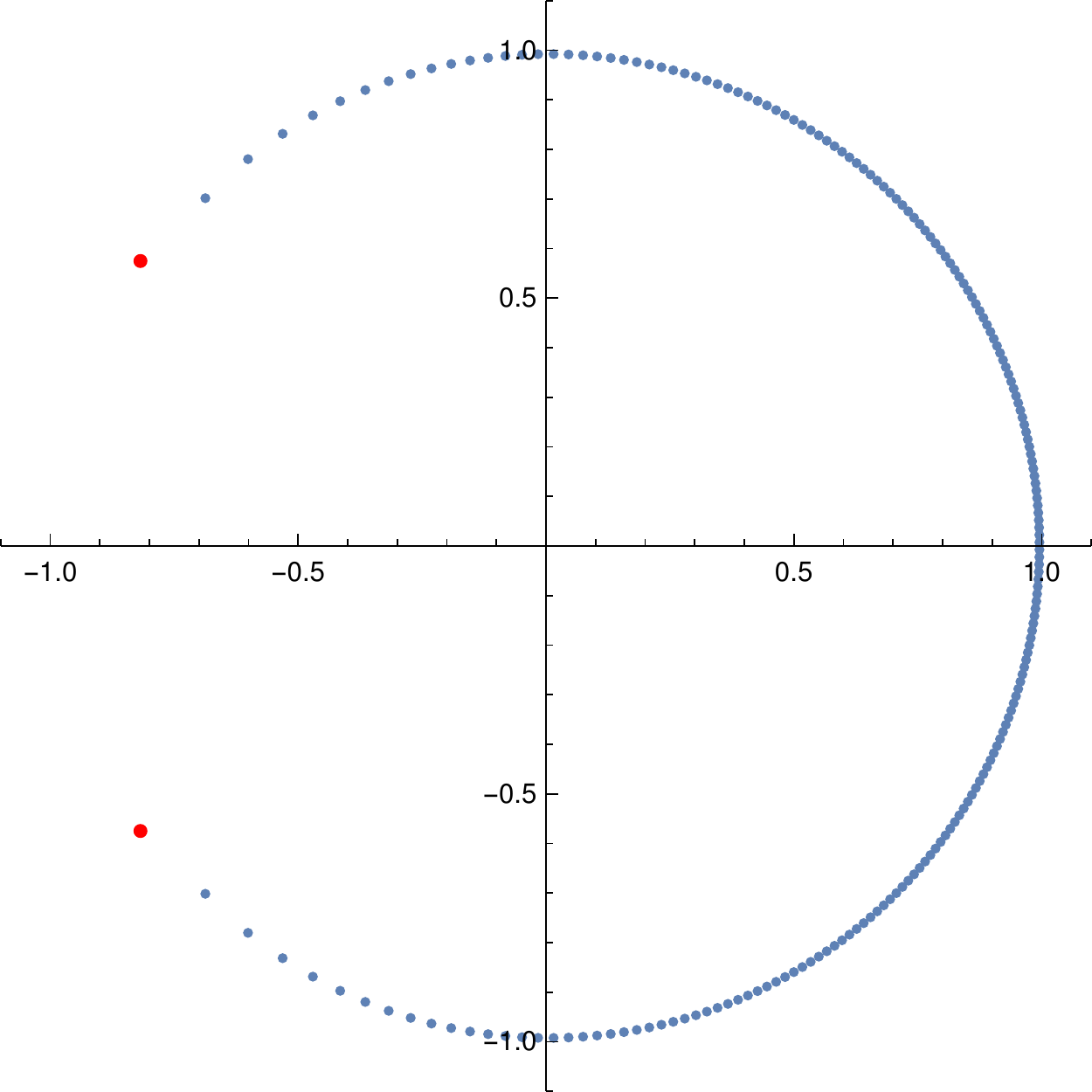}}
\hskip10mm
\subcaptionbox{$g=1.5$\label{fig:g15}}{\includegraphics[width=4.5cm]{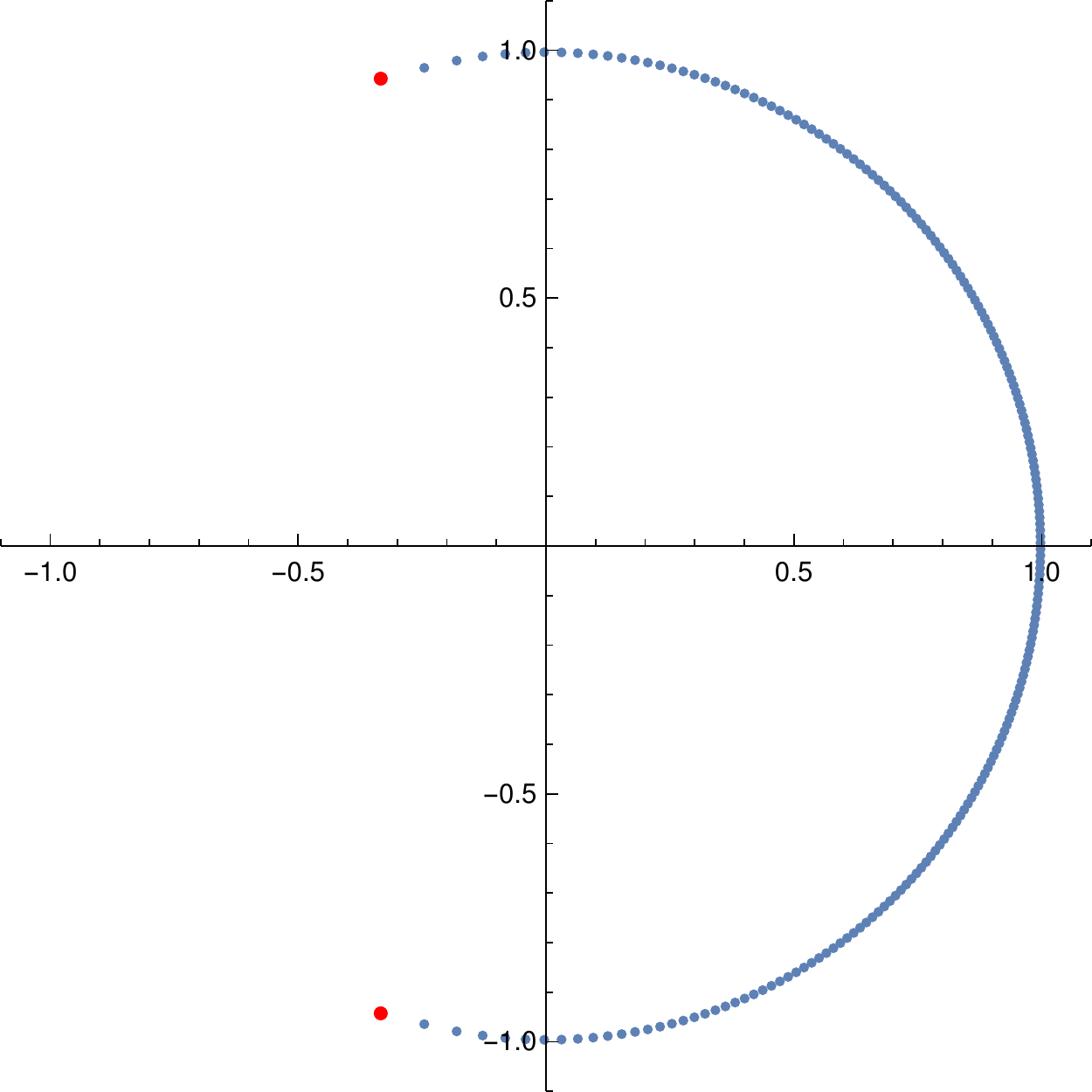}}
  \caption{
Plot 
of the eigenvalues of the matrix $M_0^{-1}M_1$
for $N=200$ at
\subref{fig:g11} $g=1.1$ and \subref{fig:g15} $g=1.5$. 
The red dots represent the end-point $e^{\pm\ri\al}$
of the cut in the large $N$ limit.
}
  \label{fig:eigen-gap}
\end{figure}

It is interesting to consider the eigenvalue distribution of 
the master field for large but finite $N$
and compare it with the known planar eigenvalue distribution of  GWW model.
First of all, the master field $M_0^{-1}M_1$ is not a unitary matrix at finite $N$,
hence it is not clear whether such a comparison is meaningful.
Nevertheless, we find numerically that in the gapped phase
the eigenvalues of $M_0^{-1}M_1$ approaches the large $N$ distribution $\rho(\th)$ in \eqref{eq:rho}
on the unit circle as $N$ becomes large
(see Fig.~\ref{fig:eigen-gap}).

\begin{figure}[htb]
\centering
\subcaptionbox{$g=0.3$\label{fig:g3}}{\includegraphics[width=5cm]{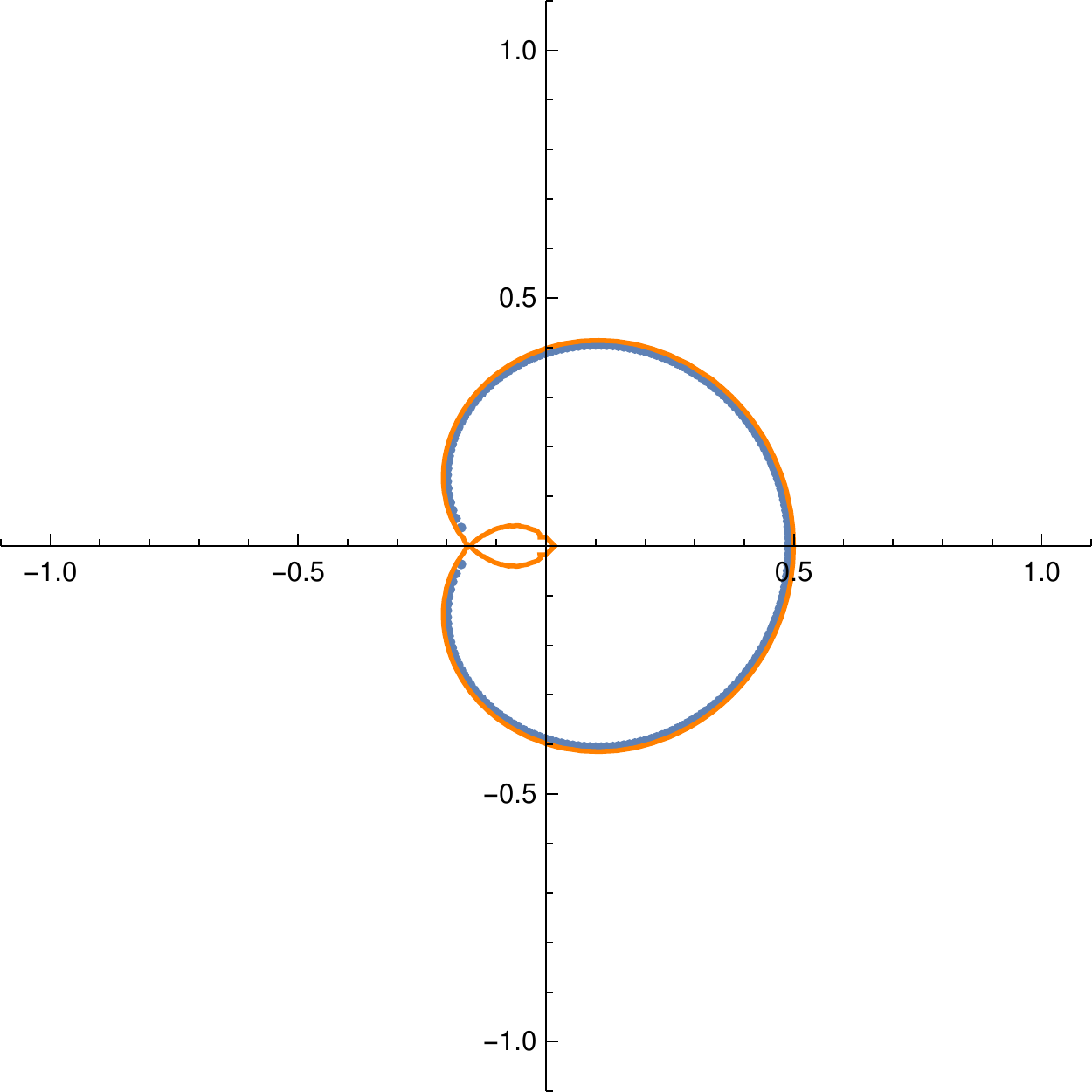}}
\hskip10mm
\subcaptionbox{$g=0.5$\label{fig:g5}}{\includegraphics[width=5cm]{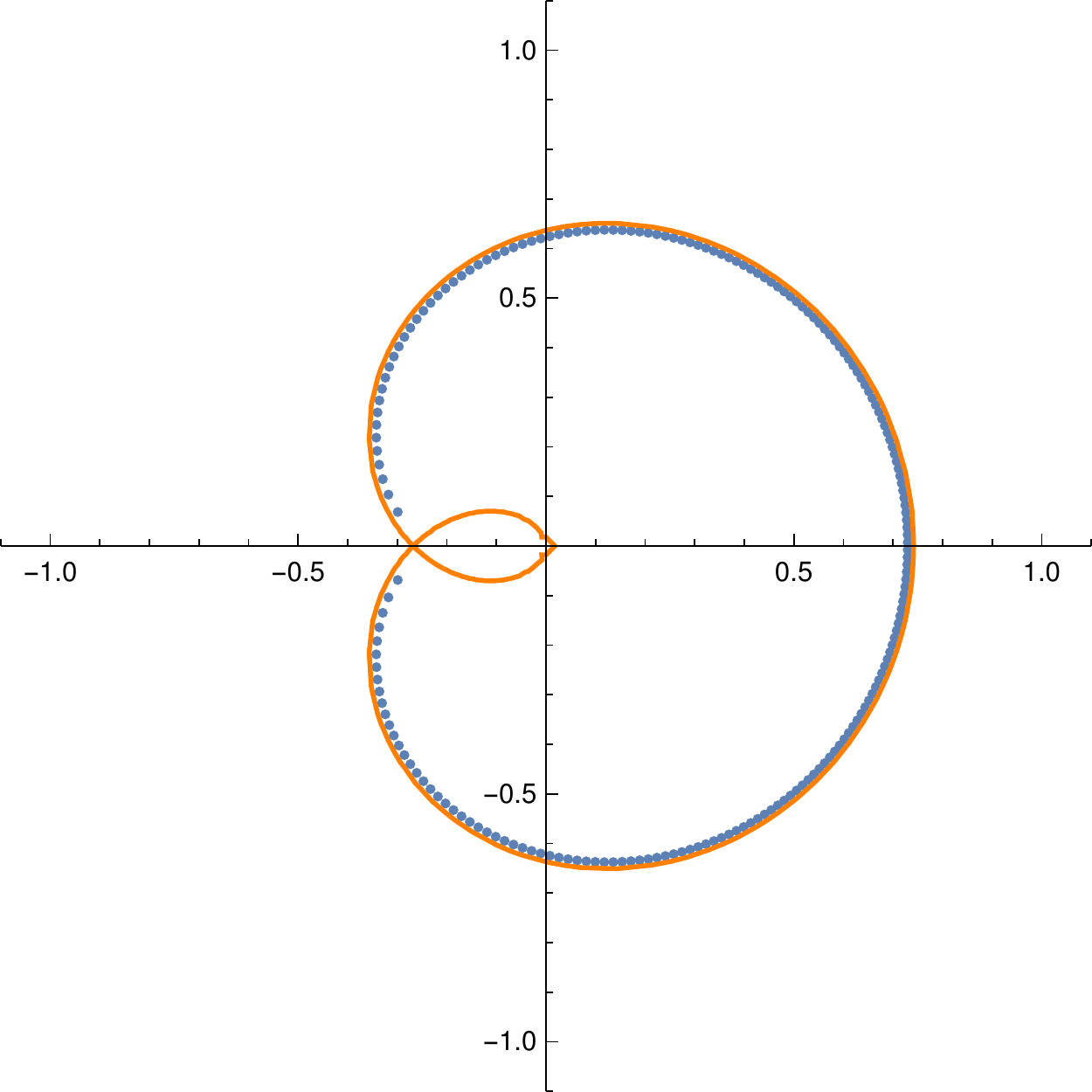}}
  \caption{
Plot 
of the eigenvalues of the matrix $M_0^{-1}M_1$
for $N=200$
at \subref{fig:g3} $g=0.3$  and  \subref{fig:g5} $g=0.5$. 
The dots are the eigenvalues of matrix $M_0^{-1}M_1$, while the orange curves
represent the equi-potential contour $\Phi(z)=-S_\text{inst}(g)$.
}
  \label{fig:eigen-ungap}
\end{figure}

On the other hand, in the ungapped phase the eigenvalues of master field are distributed inside
the unit circle (see Fig.~\ref{fig:eigen-ungap}).
Interestingly, those eigenvalues are distributed along a constant potential
contour $\Phi(z)=-S_{\text{inst}}(g)$
on the complex $z$-plane, where $S_{\text{inst}}(g)$
is the instanton action in the ungapped phase \eqref{eq:Sinst-g}
and the effective potential $\Phi(z)$
for the probe eigenvalue is given by
(see appendix \ref{app:potential})
\begin{align}
 \Phi(z)=\left\{
\begin{aligned}
-\text{Re}\Bigl[\frac{g}{2}(z-z^{-1})+\log z\Bigr],&\quad (|z|>1),\\
 \text{Re}\Bigl[\frac{g}{2}(z-z^{-1})+\log z\Bigr],&\quad (|z|<1).
\end{aligned}
\right.
\label{eq:Phi-z}
\end{align}
One can show that, in analogy with an electrostatic problem,
in the large $N$
limit the eigenvalues are distributed along the loci of constant effective potential.

\begin{figure}[tbh]
\centering\includegraphics[width=10cm]{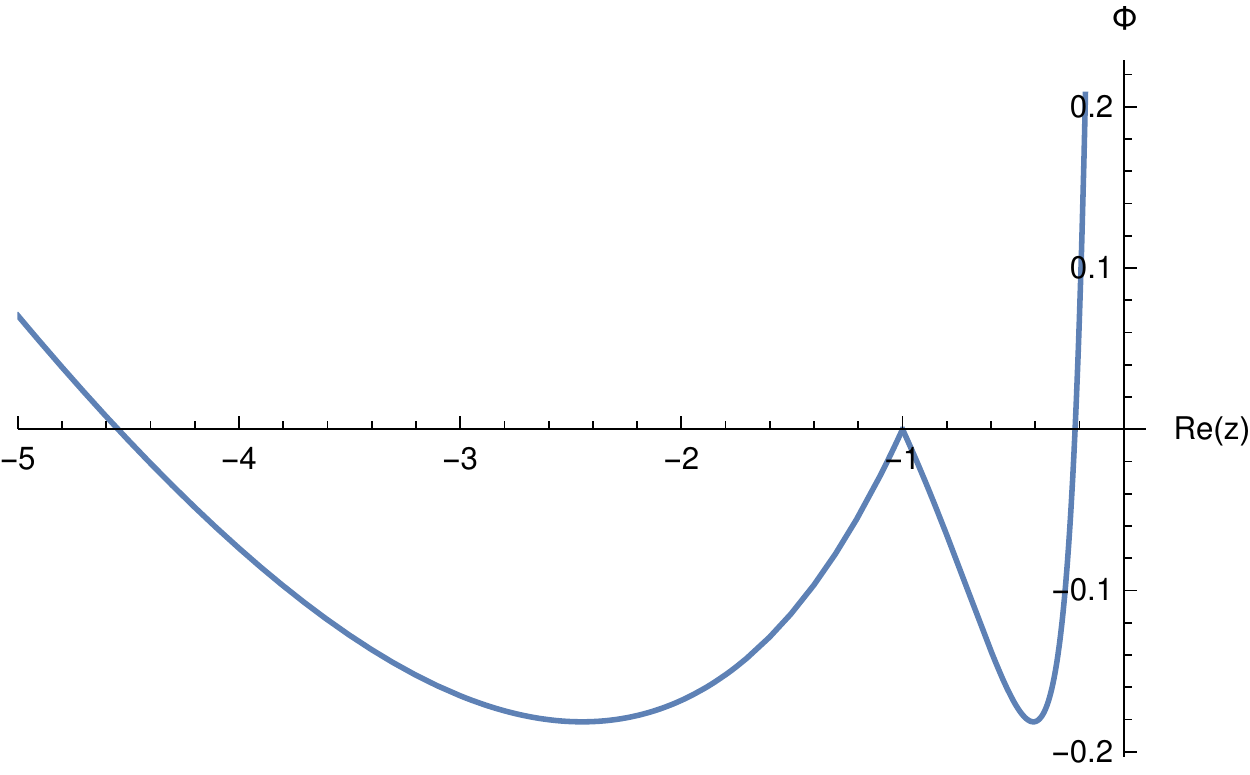}
  \caption{
Effective potential $\Phi(z)$ along the negative real $z$-axis  for $g=0.7$.
}
  \label{fig:pot}
\end{figure}

As shown in Fig.~\ref{fig:pot}, this potential has minimum at $z=z_\pm$
on the negative real $z$-axis
\begin{align}
 z_\pm=\frac{-1\pm \rt{1-g^2}}{g},
\end{align}
and the values of the potential at $z=z_\pm$ and $z=-1$
are found to be
\begin{align}
 \Phi(z_\pm)=- S_{\text{inst}}(g),\quad\Phi(-1)=0.
\label{eq:pot-value}
\end{align}
Note that the potential is constant along the unit circle
\begin{align}
 \Phi(z)=0~~~ \text{for}~~|z|=1,
\end{align}
and this is {\it higher} than the potential at $z=z_\pm$
\eqref{eq:pot-value}.\footnote{This is different from the claim in \cite{Alvarez:2016rmo}. In our notation, eq.(89) in \cite{Alvarez:2016rmo} reads
$\Phi(z_\pm)=+S_\text{inst}(g)$, but 
we believe that eq.(89) in \cite{Alvarez:2016rmo} has a sign error.}
It is tempting to identify the one-instanton correction
$\mathcal{O}(e^{-NS_{\text{inst}}(g)})$
as the effect of eigenvalue tunneling from $z=z_{-}$ to $z=-1$. 
However, it is not clear to us whether the eigenvalue distribution along the contour
$\Phi(z)=-S_{\text{inst}}(g)$ is realized as a complex  saddle 
of the GWW matrix integral.\footnote{We would like to thank 
P. Buividovich, G. Dunne, and S. Valgushev for discussion on this point.}
It would be very interesting to clarify this point further.

\section{Wilson loops in various representations \label{sec:conn}}
We can compute the expectation value of Wilson loops
in general representation exactly at finite $N$.
One can show that the expectation value of
the Wilson loop labeled by a Young diagram $\la$
is given by (see appendix \ref{app:exact})
\begin{align}
\bra\Tr_\la U \ket
=\frac{\det\Bigl[I_{\la_j+i-j}(Ng)\Bigr]}{\det\Bigl[I_{i-j}(Ng)\Bigr]}.
\label{eq:Trla-exp}
\end{align}
In this section we consider Wilson loops in ``small representations''
where the number of boxes in the corresponding Young diagram is small compared to $N$.
For small representations, 
it is convenient to use multi-trace basis rather than irreducible representations
since the connected part of multi-trace expectation value 
has a well-defined $1/N$ expansion in the gapped phase
\begin{align}
 \left\bra \prod_{i=1}^h \Tr U^{k_i}\right\ket_{\text{conn}}=\sum_{\ell=0}^\infty 
N^{2-2\ell-h} W_\ell(k_1,\cdots,k_h).
\end{align}
In the next section, we will consider Wilson loops in large representations.

For instance, using the relations
\begin{align}
 \begin{aligned}
  (\Tr U)^2&=\tableau{1 1}+\tableau{2},\\
 \Tr U\Tr U^2&=\tableau{3}-\tableau{1 1 1},\\
(\Tr U)^3&=\tableau{3}+\tableau{1 1 1}+2\tableau{2 1},
 \end{aligned}
\label{eq:multiTr-la}
\end{align}
we can compute the expectation values of the left-hand-side of \eqref{eq:multiTr-la}
by a combination of \eqref{eq:Trla-exp}.
In the gapped phase,  by numerical fitting we find the genus expansion
as 
\begin{align}
 \begin{aligned}
  \bra (\Tr U)^2\ket_{\text{conn}}&=-\frac{g-1}{g^2}
+\frac{1}{N^2}\frac{-2+3g}{8 (g-1)^2 g^2}+\cdots,\\
\bra \Tr U\Tr U^2\ket_{\text{conn}}&=
- \frac{2 (g-1) (g-2) }{g^3}
+\frac{1}{N^2}\frac{4-5g}{4 (g-1)^2 g^3}+\cdots,\\
\bra (\Tr U)^3\ket_{\text{conn}}&=
\frac{1}{N}\frac{-4+3g}{ g^3}
+\frac{1}{N^3}\frac{-8+21g-15g^2}{8 (g-1)^3 g^3}+\cdots,
 \end{aligned}
\label{eq:conn-genus}
\end{align}
while in the ungapped phase we find
the leading non-trivial instanton coefficients by numerical fitting
\begin{align}
 \begin{aligned}
  \bra (\Tr U)^2\ket_{\text{conn}}&=\frac{e^{-2NS_\text{inst}(g)}}{4\pi N}
\left[\frac{-2}{\rt{1-g^2}}+\mathcal{O}(N^{-1})\right],\\
\bra \Tr U\Tr U^2\ket_{\text{conn}}&=
\frac{e^{-2NS_\text{inst}(g)}}{4\pi N}
\left[\frac{4}{g\rt{1-g^2}}+\mathcal{O}(N^{-1})\right],\\
\bra (\Tr U)^3\ket_{\text{conn}}&=\frac{e^{-2NS_\text{inst}(g)}}{4\pi N}
\left[\frac{-4}{g}+\mathcal{O}(N^{-1})\right].
 \end{aligned}
\end{align}
It is would be interesting to compute the genus expansion analytically in the gapped phase
\eqref{eq:conn-genus} by using the relation between unitary matrix model and hermitian matrix model
as discussed in appendix \ref{app:resolvent}.

\section{Giant Wilson loops \label{sec:giant}}
In this section we consider Wilson loops in large representations, 
which are also dubbed ``Giant Wilson loops''.
In $d=4$ $\mathcal{N}=4$ SYM,
Giant Wilson loops 
are particularly interesting since they are holographically dual to some
D-brane configurations in $AdS_5\times S^5$ \cite{Drukker:2005kx,Yamaguchi:2006tq,Hartnoll:2006hr}.
In 
\cite{Grignani:2009ua,Karczmarek:2010ec,Karczmarek:2011gk},
Giant Wilson loops in unitary matrix models were studied  in the large $N$
limit. For large symmetric representation,
it was found that the there is a first order phase transition as we increase the rank of
representation.

In this section, we consider the one-loop correction to the Giant Wilson loops in GWW model 
in the $1/N$ expansion
and find a perfect match with the exact finite $N$ result. 

\subsection{Symmetric representation}
In this subsection, we consider the
Wilson loops $W_{S_k}=\bra \Tr_{S_k} U\ket$
in the $k$-th symmetric representation $S_k$.
We are interested in the regime where $k$ scales as $N$ with the ratio $x=k/N$ fixed
\begin{align}
 k,N\to\infty,\qquad x=\frac{k}{N}:~\text{fixed}.
\label{eq:giant-x}
\end{align}
It is convenient to consider the generating function of $W_{S_k}$
\begin{align}
e^{NF_S(t)} \equiv\sum_{k=0}^\infty t^kW_{S_k}=\bra\det(1-tU)^{-1}\ket,
\end{align}
and $W_{S_k}$ is 
extracted by
\begin{align}
 W_{S_k}=\oint_{t=0}\frac{dt}{2\pi\ri t^{k+1}}e^{NF_S(t)}.
\label{eq:WSk-oint}
\end{align}
In the large $N$ limit, $F_S(t)$ is given by the integral 
with the eigenvalue density $\rho(\th)$ in \eqref{eq:rho}
as a weight
\begin{align}
  F_S(t)=-\int d\th \rho(\th)\log(1-te^{\ri\th}).
\label{eq:FS-t}
\end{align}

\subsubsection*{Gapped phase}
Let us consider the generating function $F_S(t)$ \eqref{eq:FS-t}
in the gapped phase.
As shown in \cite{Karczmarek:2010ec}, the derivative
of $F_S(t)$ in the planar limit can be written in a closed form
\begin{align}
\begin{aligned}
 t\del_t F_S=\int d\th \rho(\th)\frac{te^{\ri\th}}{1-te^{\ri\th}}
=-\hf +\frac{g(t+1)}{4t}\Biggl[t-1+\rt{(t-1)^2+\frac{4t}{g}}\Biggr]. 
\end{aligned}
\end{align}
In the limit \eqref{eq:giant-x}, the integral \eqref{eq:WSk-oint}
can be evaluated by the saddle point approximation, where 
the saddle point equation reads
\begin{align}
 t\del_t F_S=x,
\end{align}
and the solution of saddle point equation is given by
\begin{align}
 t_*=\frac{(1+2x)^2-2g+ (1+2x)\rt{(1+2x)^2+4g(g-1)}}{4g(1+x)}.
\end{align}
The saddle point value is evaluated as
\begin{align}
 \begin{aligned}
  F_S(t_*)-x\log t_*&=
\hf\Bigl(1-2g+\rt{(1+2x)^2+4g(g-1)}\Bigr)\\
&+\log\frac{1+2x+2g-\rt{(1+2x)^2+4g(g-1)}}{2}-x\log t_*.
 \end{aligned}
\end{align}
One can also compute the one-loop correction
from the Gaussian fluctuation around the saddle point.
At this order we do not need the genus-one correction to $\rho(\th)$.
Finally, we find
\begin{align}
 \log W_{S_k}=N \Bigl[F_S(t_*)-x \log t_*\Bigr]
-\hf \log\Bigl[2\pi N F_S''(t_*)\Bigr],
\label{eq:gap-sym-oneloop}
\end{align}
where $F_S''(t_*)$ denotes the second derivative of $F_S$ with respect to $\log t$
\begin{align}
 F_S''(t_*)=\frac{(1+2x)^2+2(g-1)-\rt{(1+2x)^2+4g(g-1)}}{2(1+2x)^2+8(g-1)}\rt{(1+2x)^2+4g(g-1)}.
\end{align}
\begin{figure}[tbh]
\centering\includegraphics[width=8cm]{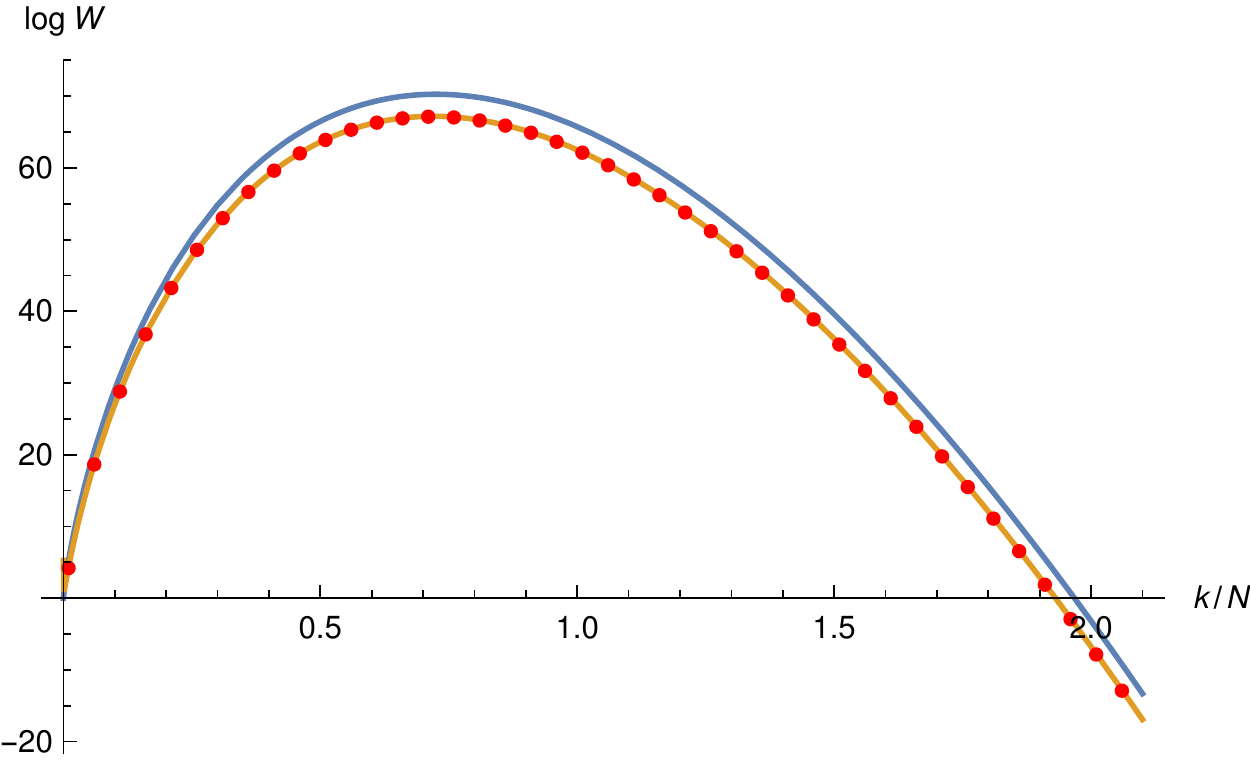}
  \caption{
Plot of $\log W_{S_k}$ in the gapped phase ($g=1.5, N=100$).
The red dots are the exact values, while
the blue curve and the orange curve represent
the leading term and the leading$+$one-loop correction in \eqref{eq:gap-sym-oneloop}, respectively.
One can see that the one-loop correction
improves the matching with the exact result.}
  \label{fig:sym-gap}
\end{figure}

In Fig.~\ref{fig:sym-gap}, we show the plot of $\log W_{S_k}$ as a function of $x=k/N$
for $g=1.5$. One can see that including the one-loop correction (i.e. the second 
term in \eqref{eq:gap-sym-oneloop}) improves the matching with the exact value of $\log W_{S_k}$
at finite $N$. 

\subsubsection*{Ungapped phase} 
\begin{figure}[tbh]
\centering\includegraphics[width=8cm]{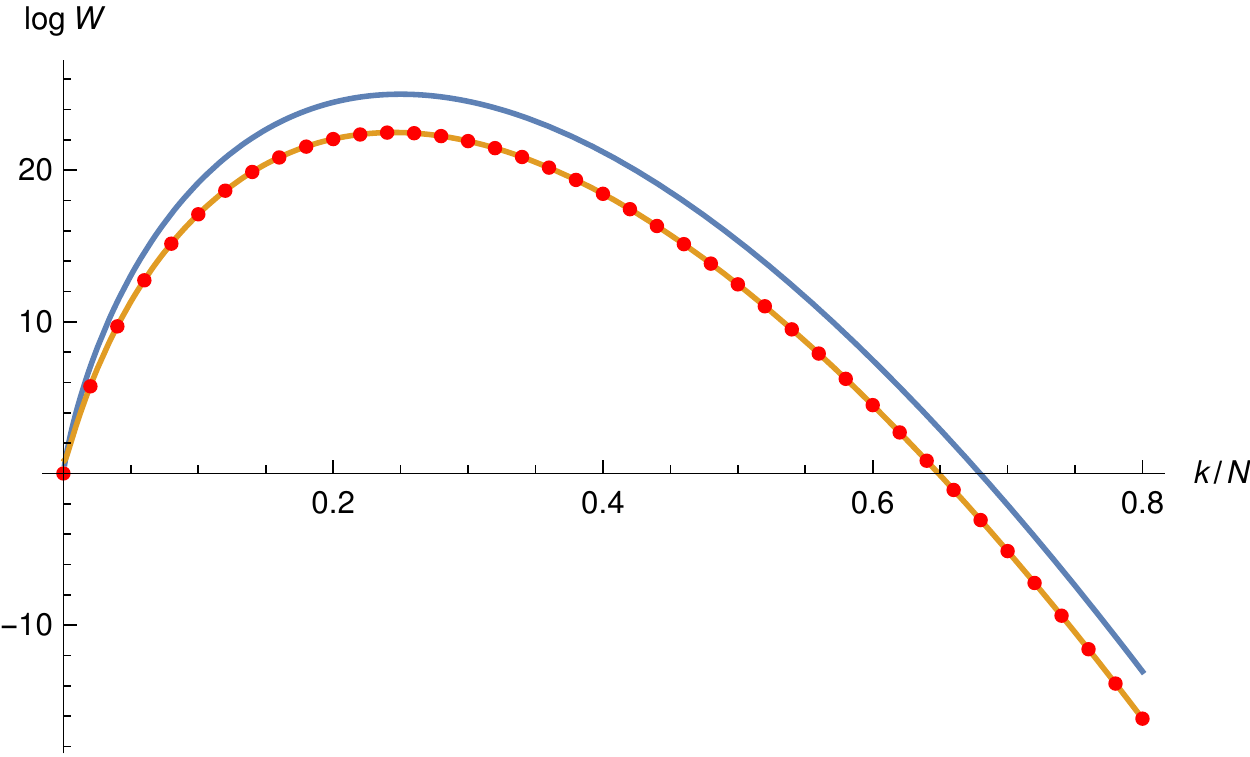}
  \caption{
Plot of $\log W_{S_k}$ in the ungapped phase ($g=0.5, N=100$).
The red dots are the exact values, while
the blue curve and the orange curve represent
the leading term and the leading$+$one-loop correction in \eqref{eq:GWW-Sk-oneloop}, respectively.
Again, one can see that the inclusion of the one-loop correction
improves the matching.}
  \label{fig:WSk-ungap}
\end{figure}

In the ungapped phase, 
$W_{S_k}$ is dominated by the $\bra \Tr U\ket^k$
term since higher traces $\Tr U^m~(m\geq2)$ are exponentially suppressed
in the large $N$ limit
\cite{Grignani:2009ua}
\begin{align}
 W_{S_k}\approx \frac{1}{k!}\bra \Tr U\ket^k
\approx \frac{1}{k!}\left(\frac{Ng}{2}\right)^k.
\end{align}
Using the Stirling's formula
\begin{align}
 k!\approx \rt{2\pi Nx}\left(\frac{Nx}{e}\right)^{Nx},
\end{align}
we find
\begin{align}
 \log W_{S_k}=Nx\log\Bigl(\frac{eg}{2x}\Bigr)-\hf\log(2\pi Nx).
\label{eq:GWW-Sk-oneloop}
\end{align}
The second term can be thought of as the ``one-loop'' correction
to the result in \cite{Grignani:2009ua}.
Again, as shown in Fig.~\ref{fig:WSk-ungap},
the one-loop correction 
improves the matching with the exact result at finite $N$.

\subsection{Anti-symmetric representation}
In this section we consider the Wilson loops $W_{A_k}=\bra \Tr_{A_k}U\ket$
of GWW model
in the $k$-th anti-symmetric representation $A_k$ in the limit \eqref{eq:giant-x}.
As in the case of symmetric representation, it is convenient to consider the generating function of
$W_{A_k}$
\begin{align}
 e^{NF_A(t)}\equiv\sum_{k=0}^N t^k W_{A_k}=\bra \det(1+tU)\ket.
\end{align}
In the large $N$ limit, $F_A(t)$ is given by an integral with weight $\rho(\th)$
\begin{align}
 F_A(t)=\int d\th \rho(\th)\log(1+te^{\ri\th})
\end{align}
and the $W_{A_k}$ is given by
\begin{align}
 W_{A_k}=\oint\frac{dt}{2\pi\ri t^{k+1}}e^{NF_A(t)}.
\label{eq:WAk-oint}
\end{align} 
\subsubsection*{Gapped phase}
\begin{figure}[tbh]
\centering\includegraphics[width=8cm]{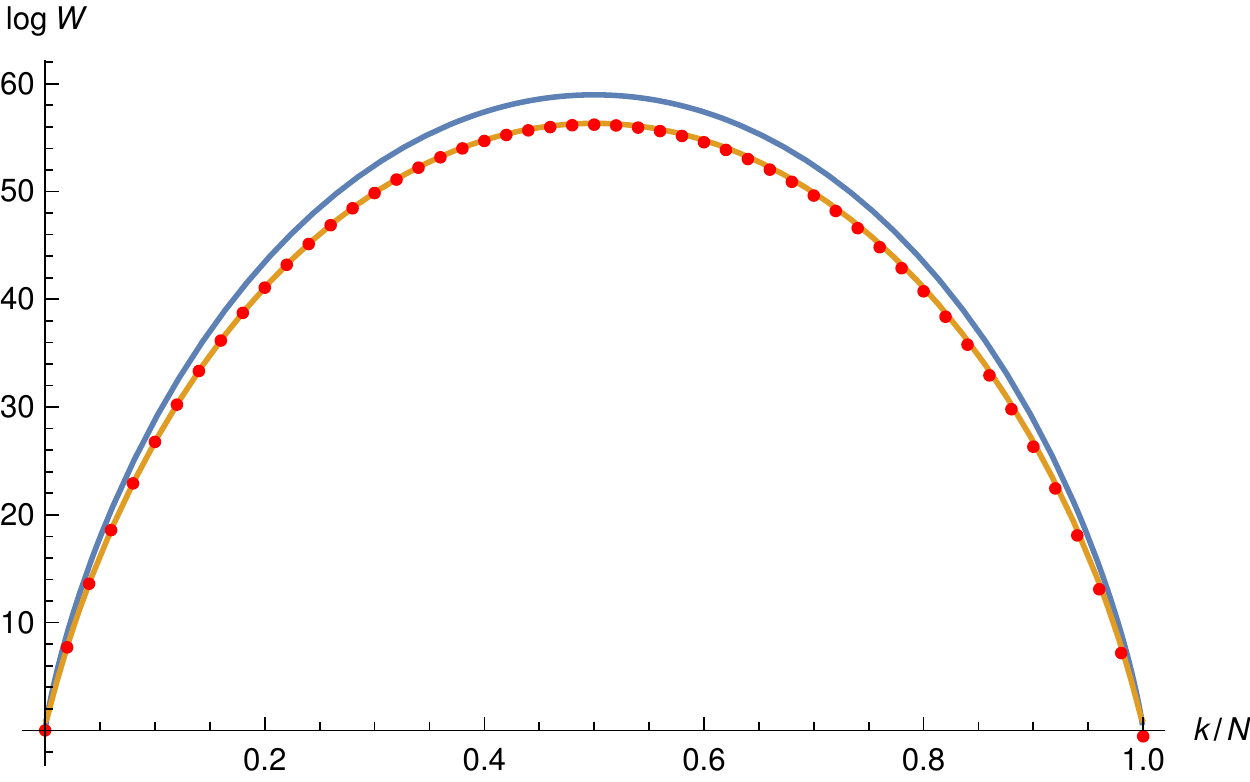}
  \caption{
Plot of the log of Wilson loop in the anti-symmetric
representation $\log W_{A_k}$ as a function of $k/N$
for $g=1.5, N=100$.
The red dots are the exact values while 
the blue curve and the orange curve are
the leading term and the leading$+$one-loop correction in \eqref{eq:GWW-asym-oneloop}, respectively.
One can clearly see that the inclusion of the one-loop correction
improves the matching with the exact value.
}
  \label{fig:WAk-gap}
\end{figure}

Let us consider 
$W_{A_k}$ 
in the gapped phase.
Again,
in the limit \eqref{eq:giant-x} the integral \eqref{eq:WAk-oint}
can be evaluated by the saddle point approximation.
The saddle point equation is
\begin{align}
 t\del_t F_A=x,
\end{align}
where the left-hand-side is computed as
\begin{align}
\begin{aligned}
 t\del_t F_A
=\int d\th \rho(\th)\frac{te^{\ri\th}}{1+te^{\ri\th}}
=\hf+\frac{g(t-1)}{4t}\Biggl[t+1-\rt{(t+1)^2-\frac{4t}{g}}\Biggr].
\end{aligned}
\end{align}
There are two solutions of saddle point equation,
but the solution corresponding to the dominant saddle 
turns out to be \cite{Karczmarek:2011gk}
\begin{align}
 t_*=\frac{2g-(1-2x)^2- (1-2x)\rt{(1-2x)^2-4g(1-g)}}{4g(1-x)},
\end{align}
and the saddle point value is
\begin{align}
 \begin{aligned}
  F_A(t_*)-x\log t_*&=\hf\Bigl(2g-1-\rt{(1-2x)^2-4g(1-g)}\Bigr)\\
&+\hf\log\frac{2g-1+\rt{(1-2x)^2-4g(1-g)}}{4gx(1-x)}\\
&-\frac{x}{2}\log t_{*}(x)-\frac{1-x}{2}\log t_{*}(1-x).
 \end{aligned}
\label{eq:saddle-value-Ak}
\end{align}
Note that \eqref{eq:saddle-value-Ak} is symmetric under the exchange
$x\lrya 1-x$.
One can also compute the one-loop correction
by performing the Gaussian integral around the saddle point
\begin{align}
 \log W_{A_k}=N \Bigl[ F_A(t_*)-x\log t_*\Bigr]
-\hf \log\Bigl[2\pi N F_A''(t_*)\Bigr],
\label{eq:GWW-asym-oneloop}
\end{align}
where
\begin{align}
 F_A''(t_*)=\frac{(1-2x)^2+2(g-1)+\rt{(1-2x)^2-4g(1-g)}}{2(1-2x)^2+8(g-1)}\rt{(1-2x)^2-4g(1-g)}.
\end{align}
As one can see from Fig.~\ref{fig:WAk-gap},
matching with the exact value at finite $N$ is improved by including the one-loop correction.

\subsubsection*{Ungapped phase}
\begin{figure}[tbh]
\centering\includegraphics[width=8cm]{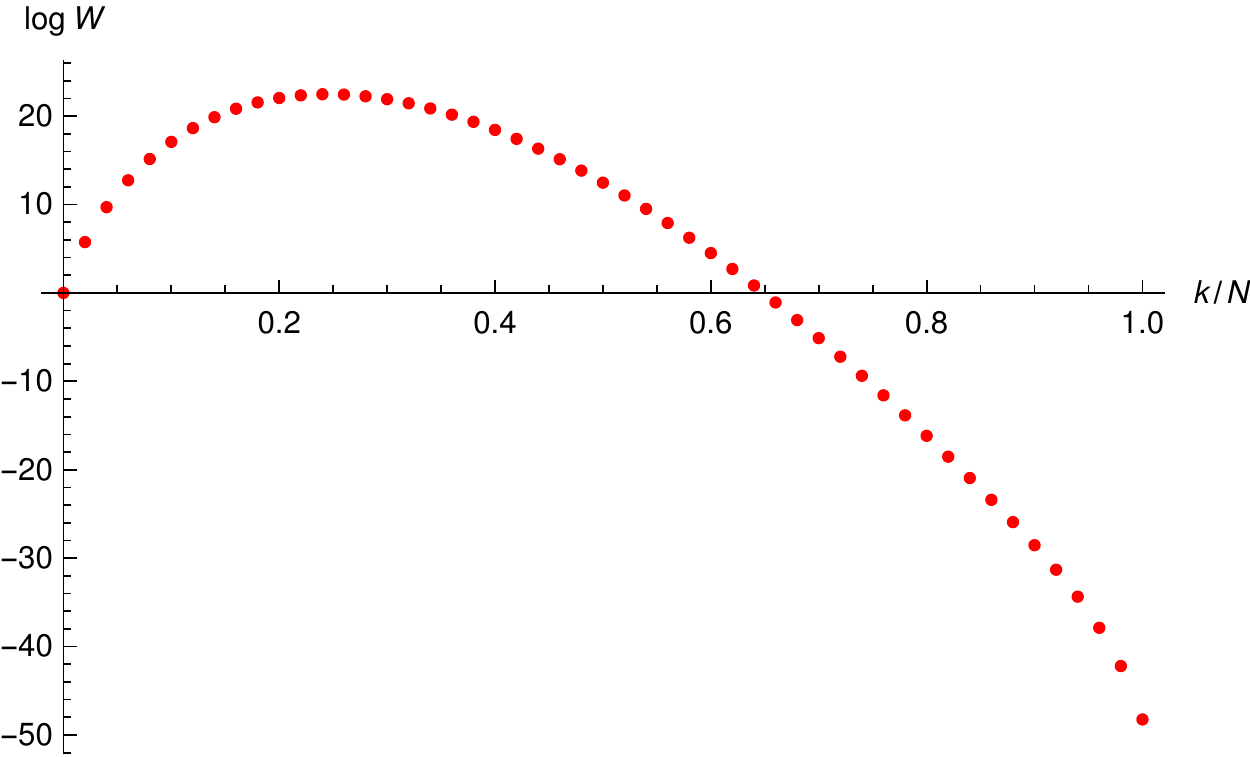}
  \caption{
Plot of $\log W_{A_k}$ in the ungapped phase ($g=0.5, N=100$).
We do not have a symmetry $k\lrya N-k$ in the $U(N)$ theory.
}
  \label{fig:Ak-low}
\end{figure}

In \cite{Karczmarek:2010ec},
it was found that in the ungapped phase the
symmetry $k\to N-k$ of $W_{A_k}$ is realized by a first order
phase transition for the model with gauge group $SU(N)$.
In our case of $U(N)$ matrix model,
there is no such symmetry at finite $N$,
although we have an approximate symmetry $k\to N-k$ in the gapped phase
in the large $N$ limit (see Fig.~\ref{fig:WAk-gap}).
As shown in Fig.~\ref{fig:Ak-low},
we indeed find that the $W_{A_k}$
is not symmetric under $k\to N-k$ in the ungapped phase.
It would be interesting to find the exact form of $W_{A_k}$ for $SU(N)$
theory at finite $N$
and confirm the result of \cite{Karczmarek:2010ec}.

\subsection{Rectangular Young diagram}
\begin{figure}[htb]
\centering
\subcaptionbox{$n=2$\label{fig:n2}}{\includegraphics[width=6cm]{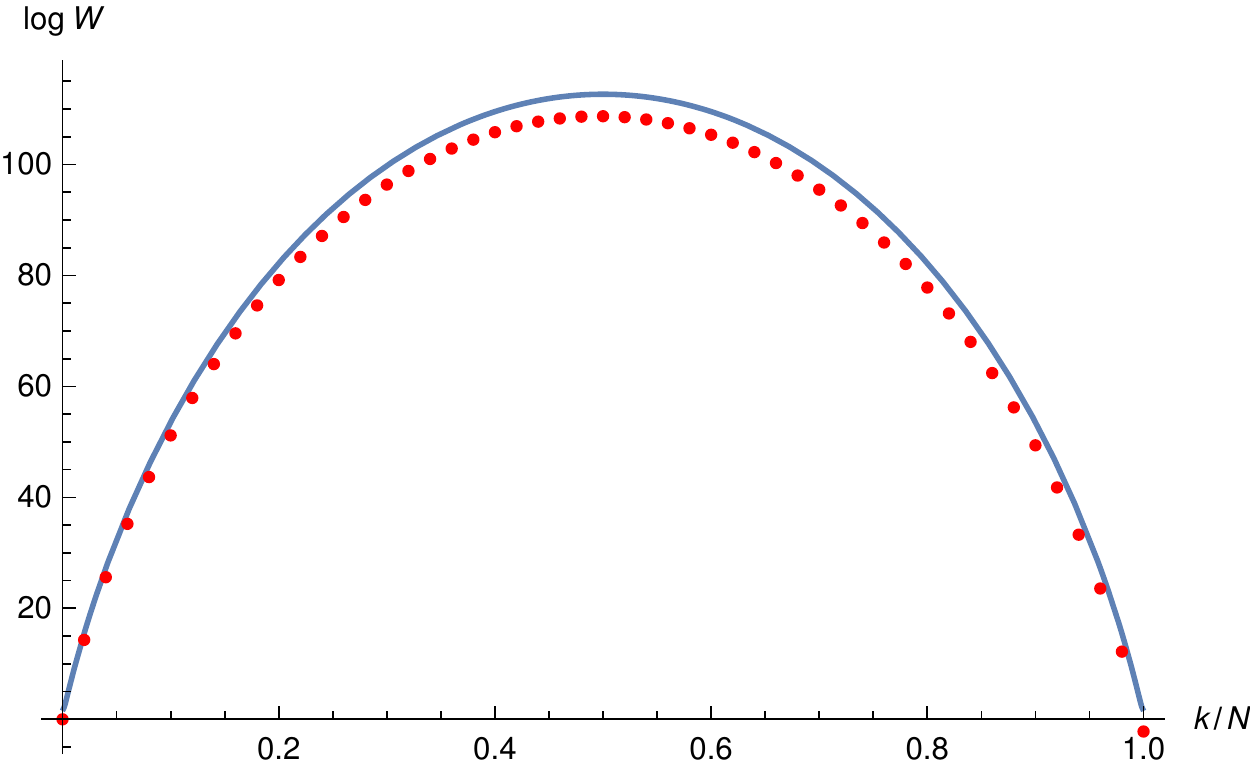}}
\hskip15mm
\subcaptionbox{$n=3$\label{fig:n3}}{\includegraphics[width=6cm]{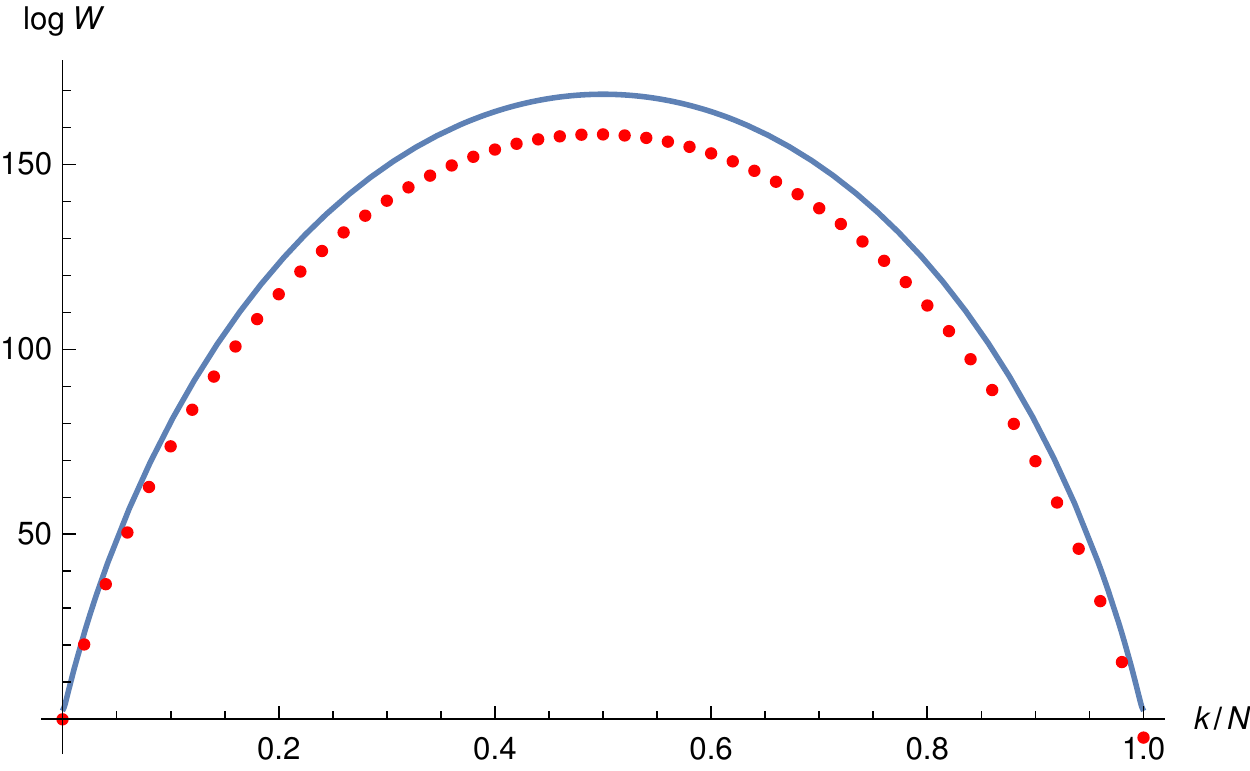}}
  \caption{
Plot 
of the $\log W_\la$ in  the representation
$\la=[n^k]$ for
\subref{fig:n2} $n=2$ and \subref{fig:n3} $n=3$ as a function of $k/N$ with $N=100$. 
The red dots are the exact values of $\log W_{[n^k]}$, while the solid curves
represent $n \log W_{A_k}$.
}
  \label{fig:bind}
\end{figure}

In the case of  $\mathcal{N}=4$ SYM,
Giant Wilson loops in the representation associated with the
rectangular Young diagram are holographically dual to multiple D5 or D3-branes
\cite{Gomis:2006sb,Gomis:2006im}.
In the GWW model we also expect that Giant Wilson loops 
associated with rectangular Young diagram
have a simple relation to the (anti-)symmetric Wilson loops.
In particular, we expect that the Wilson loop $W_\la$
for the Young diagram $\la=[n^k]$
is related to the $n$-th power of the anti-symmetric Wilson loop $W_{A_k}=W_{[1^k]}$ 
\begin{align}
 W_{[n^k]}\sim (W_{[1^k]})^n.
\label{eq:rec-young}
\end{align}
However, we find numerically that 
the relation \eqref{eq:rec-young} holds only approximately and
in general we have an inequality (see Fig.~\ref{fig:bind})
\begin{align}
 \log W_{[n^k]}<n\log W_{[1^k]}.
\end{align}
The difference $n\log W_{[1^k]}-\log W_{[n^k]}$ might be 
physically interpreted as the binding energy
between multiple Giant loops in GWW model.

\section{Adjoint model \label{sec:adj}}
In this section we consider a unitary matrix model with double trace 
interaction
\begin{align}
 \mathcal{Z}(N,a)=\int_{U(N)} dU \exp\Bigl(a\Tr U\Tr U^\dagger\Bigr).
\end{align}
We call this model the ``adjoint model'' since $\Tr U\Tr U^\dagger=\Tr_{\text{adj}}U$
is the trace in the adjoint representation of $U(N)$.
This model can be thought of as a truncation of 
the thermal partition function of free $\mathcal{N}=4$ SYM
on $S^3\times S^1$
\footnote{If we turn on the interaction, 
the thermal partition function
can be described by an effective model
with one more parameter $b$ \cite{Aharony:2003sx,AlvarezGaume:2005fv}
\begin{align}
 \mathcal{Z}(N,a,b)=\int_{U(N)} dU \exp\Biggl(a|\Tr U|^2+\frac{b}{N^2}|\Tr U|^4\Biggr).
\end{align}
In this paper we only consider the special case $b=0$.
},
and it is known that this model exhibits a Hagedorn/deconfinement transition
at $a=1$. In the low temperature regime $(a<1)$ this model is in the confined
phase and the free energy is $\mathcal{O}(N^0)$ while
in the high temperature regime $(a>1)$ 
this model is in the deconfined
phase and the free energy is $\mathcal{O}(N^2)$.

As discussed in \cite{Liu:2004vy}, the partition function of the adjoint model $\mathcal{Z}(N,a)$
and that of the GWW model $Z(N,g)$
are related by a certain integral transformation
\begin{align}
 \mathcal{Z}(N,a)=\frac{N^2}{2a}\int_0^\infty g dg e^{-\frac{N^2g^2}{4a}}Z(N,g).
\label{eq:Zadj-int}
\end{align}
Using the exact result of $Z(N,g)$ in \eqref{eq:Z-exact},
one can compute $\mathcal{Z}(N,a)$
at finite $N$ by evaluating the integral \eqref{eq:Zadj-int} numerically.

\subsection*{Free energy of the adjoint model}
Now let us consider the free energy of adjoint model.
As emphasized in \cite{Liu:2004vy},
the partition function of the adjoint model in \eqref{eq:Zadj-int}
can be naturally written as a sum of two contributions
\begin{align}
  \mathcal{Z}(N,a)= \mathcal{Z}_{\text{th-AdS}}(N,a)+\mathcal{Z}_{\text{BBH}}(N,a),
\end{align}
where
\begin{align}
\begin{aligned}
 \mathcal{Z}_{\text{th-AdS}}(N,a)&=
\frac{N^2}{2a}\int_0^1 g dg e^{-\frac{N^2g^2}{4a}}Z(N,g),\\
\mathcal{Z}_{\text{BBH}}(N,a)&=
\frac{N^2}{2a}\int_1^\infty g dg e^{-\frac{N^2g^2}{4a}}Z(N,g),
\end{aligned}
\end{align}
and $\mathcal{Z}_{\text{th-AdS}}(N,a)$ and 
$\mathcal{Z}_{\text{BBH}}(N,a)$ are interpreted as 
the contributions of the thermal AdS and the AdS-Schwarzchild black hole (big black hole),
respectively. 
On the bulk gravity side, the deconfinement transition at $a=1$
corresponds to the Hawking-Page transition
where  the thermal $AdS$ and the big black hole exchange dominance \cite{Witten:1998zw}.

\begin{figure}[htb]
\centering
\subcaptionbox{$1/2<a<2$\label{fig:whole-a}}{\includegraphics[width=4.5cm]{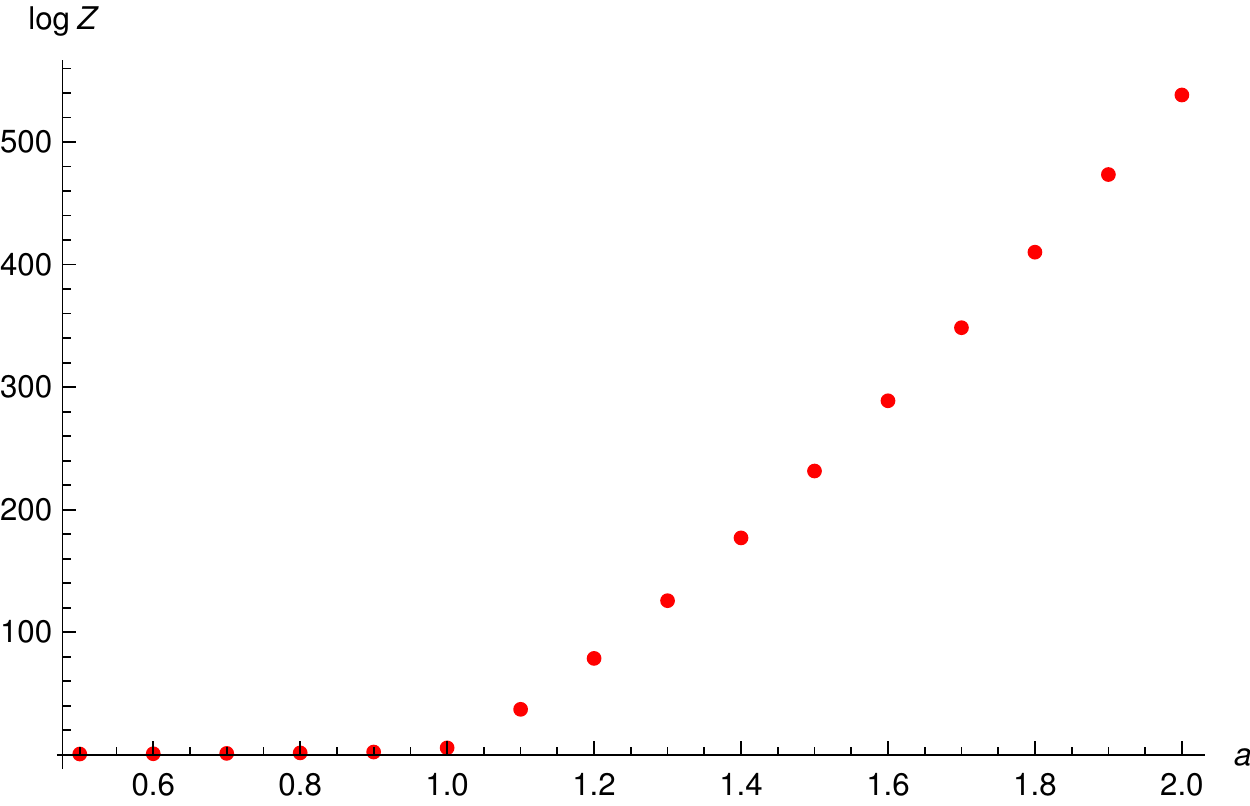}}
\hskip8mm
\subcaptionbox{$1/2<a<1$\label{fig:alow}}{\includegraphics[width=4.5cm]{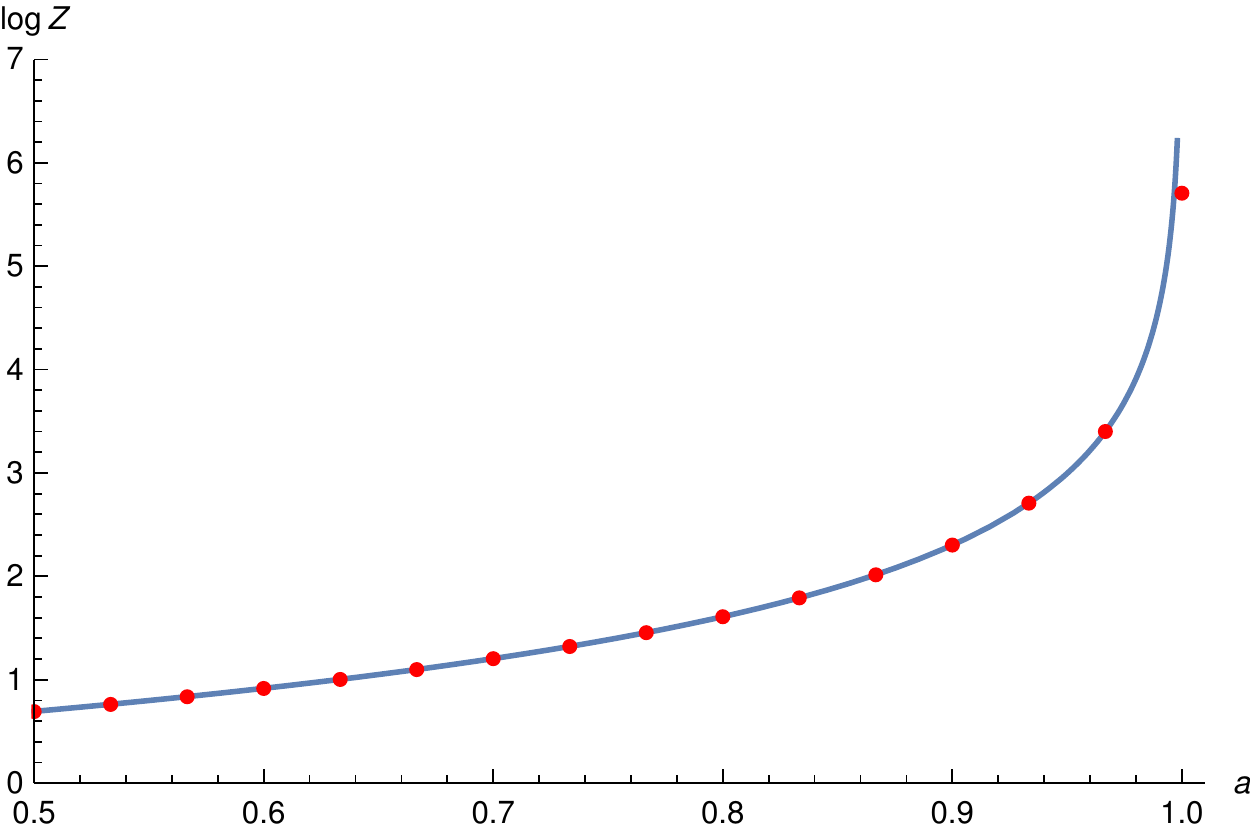}}
\hskip8mm
\subcaptionbox{$1<a<2$\label{fig:ahigh}}{\includegraphics[width=4.5cm]{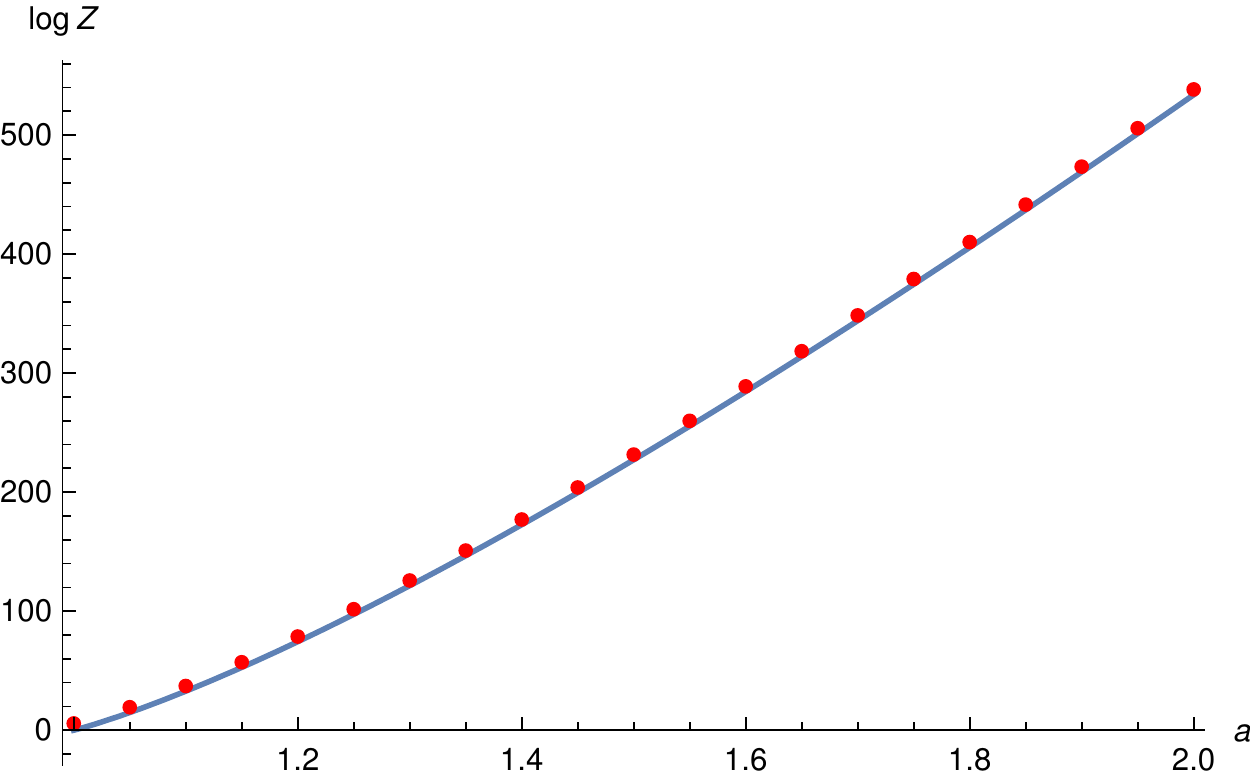}}
  \caption{
Plot of free energy $\log\mathcal{Z}(N,a)$  for $N=30$ in the range
$1/2<a<2$. In \subref{fig:whole-a}, we show the plot in the whole region 
$1/2<a<2$, while in \subref{fig:alow} and \subref{fig:ahigh} we magnify the region $a<1$ and  $a>1$, respectively. 
The red dots are the numerical value of the free
energy. The solid curves in \subref{fig:alow} and \subref{fig:ahigh}
represent $-\log(1-a)$ and  $N^2\mathcal{F}_0(a)$ in \eqref{eq:lead-free-a}, respectively.
}
  \label{fig:free-alh}
\end{figure}

In the large $N$ limit, the partition function of GWW model
$Z(N,g)$ can be replaced by its
planer limit $Z(N,g)\approx e^{N^2F_0(g)}$ in \eqref{eq:F0-GW},
and it turns out that the 
$g$-integral \eqref{eq:Zadj-int}
is dominated by $\mathcal{Z}_{\text{th-AdS}}(N,a)$
in the confined phase $(a<1)$ and by $\mathcal{Z}_{\text{BBH}}(N,a)$
in the deconfined phase $(a>1)$.
The free energy of the adjoint model is computed as
\begin{align}
\log\mathcal{Z}(N,a)\approx \left\{\begin{aligned}
 &-\log(1-a),&\qquad (a<1),\\
&N^2\mathcal{F}_0(a),&\qquad (a>1),
 \end{aligned}\right.
\label{eq:lead-free-a}
\end{align}
where the genus-zero free energy $\mathcal{F}_0(a)$ in the deconfined phase
is given by
\begin{align}
 \mathcal{F}_0(a)=-\frac{g_*^2}{4a}+g_*-\hf\log g_*-\frac{3}{4}
\end{align}
with $g_*$ being the saddle point value of $g$
\begin{align}
 g_*=a+\rt{a(a-1)}.
\end{align}
As shown in Fig.~\ref{fig:free-alh},
the free energy for $N=30$ evaluated numerically by \eqref{eq:Zadj-int}
nicely reproduces the analytic result \eqref{eq:lead-free-a} at the leading order in the large $N$ expansion.
One can proceed to study subleading corrections in the large $N$ expansion.
In the deconfined phase $a>1$, the free energy has a standard genus expansion
\begin{align}
 \log\mathcal{Z}(N,a)=\sum_{\ell=0}^\infty N^{2-2\ell}\mathcal{F}_\ell(a).
\label{eq:Fa-genus}
\end{align}
In particular, the genus-one free energy is given by
\begin{align}
 \mathcal{F}_1(a)&=F_1(g_*)+\log\left[\frac{\rt{\pi}Ng_*}{a\rt{1/a-1/g_*^2}}\right],
\label{eq:F1-amodel}
\end{align}
where $F_1(g)$ is the genus-one free energy of GWW model in \eqref{eq:Fell-anal}. The second term 
of \eqref{eq:F1-amodel} comes from the Gaussian integral around the saddle point $g=g_*$.
\begin{figure}[tbh]
\centering\includegraphics[width=8cm]{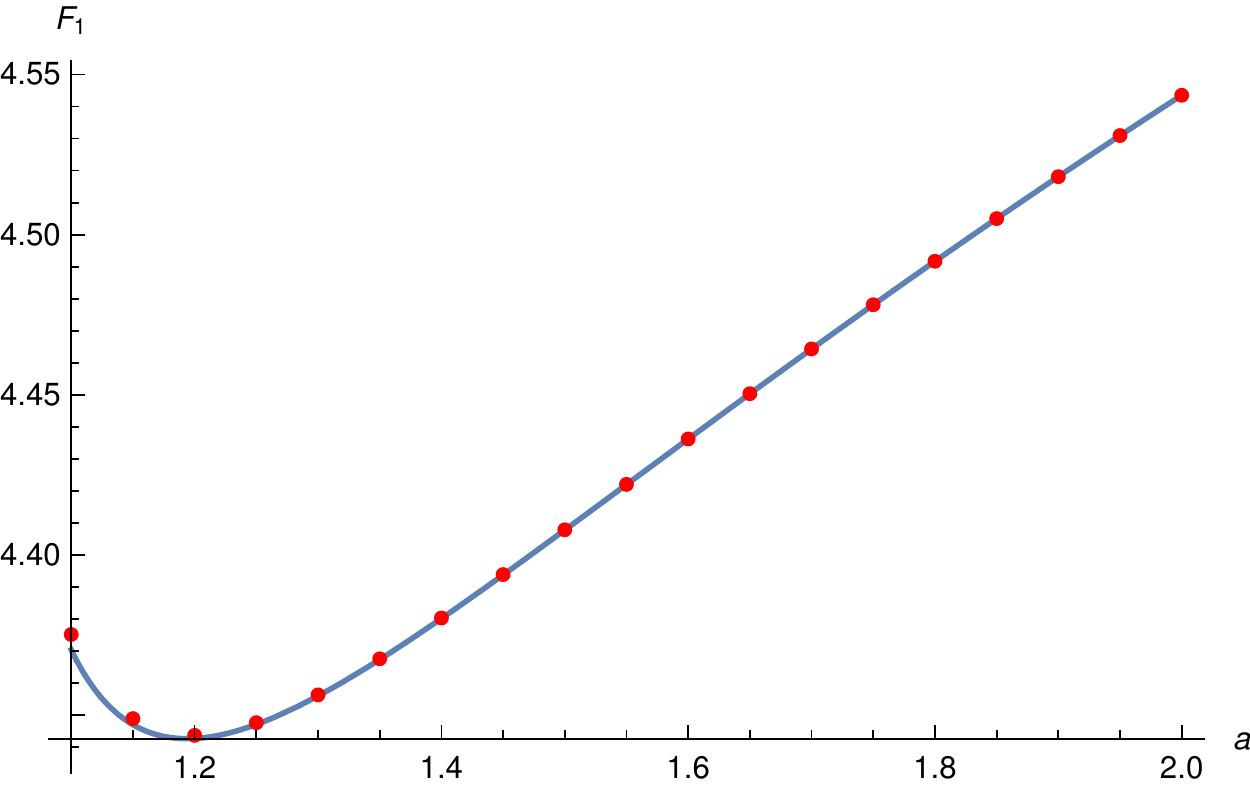}
  \caption{
Plot of the genus-one free energy $\mathcal{F}_1(a)$ in the deconfined phase $a>1$.
The red dots are the numerical value of $\log \mathcal{Z}(N,a)-N^2\mathcal{F}_0(a)$
for $N=30$,
while the solid curve is the plot of  the analytic form of $\mathcal{F}_1(a)$ in \eqref{eq:F1-amodel}.
}
  \label{fig:free-aloop}
\end{figure}
As one can see from Fig.~\ref{fig:free-aloop}, 
after subtracting the genus-zero part the free energy for $N=30$ exhibits a nice agreement with
the analytic form of one-loop correction \eqref{eq:F1-amodel}.
It would be interesting study the higher genus corrections $\mathcal{F}_\ell(a)$
in \eqref{eq:Fa-genus}. 

In the confined phase, it is expected that there is a non-perturbative correction
to the leading result \eqref{eq:lead-free-a}
and the apparent singularity at the transition point $a=1$
is smoothed out \cite{Liu:2004vy}.
It would be very interesting to study such non-perturbative corrections in detail
and find a possible bulk string theory interpretation.
We leave this as an interesting future problem.

\subsection*{Winding loops in the adjoint model}
The expectation value of Wilson loops in the adjoint model\footnote{In the context
of $\mathcal{N}=4$ SYM
on $S^3\times S^1$,  Wilson loops in the adjoint model
are interpreted as Polyakov loops
wrapping the thermal $S^1$.}
can also be written as a certain integral transform of that of the GWW model.
For general operator $\mathcal{O}$, its
expectation value  $\bra \mathcal{O}\ket_{a}$ in the adjoint model
is given by\footnote{This can be thought of as a disorder average over the random coupling $g$, which is reminiscent of the Sachdev-Ye-Kitaev model \cite{SY,Kitaev}.}
\begin{align}
 \bra \mathcal{O}\ket_{a}=\frac{\int dU \mathcal{O} \exp(a\Tr U\Tr U^\dagger)}
{\int dU \exp(a\Tr U\Tr U^\dagger)}=
\frac{\int_0^\infty dg g e^{-\frac{N^2g^2}{4a}}\int dU \mathcal{O} 
\exp[\frac{Ng}{2}\Tr (U+ U^\dagger)]}
{\int_0^\infty dg g e^{-\frac{N^2g^2}{4a}}\int dU \exp[\frac{Ng}{2}\Tr (U+ U^\dagger)]}.
\label{eq:veva-int}
\end{align}

\begin{figure}[htb]
\centering
\subcaptionbox{$u_1$\label{fig:F1}}{\includegraphics[width=4.5cm]{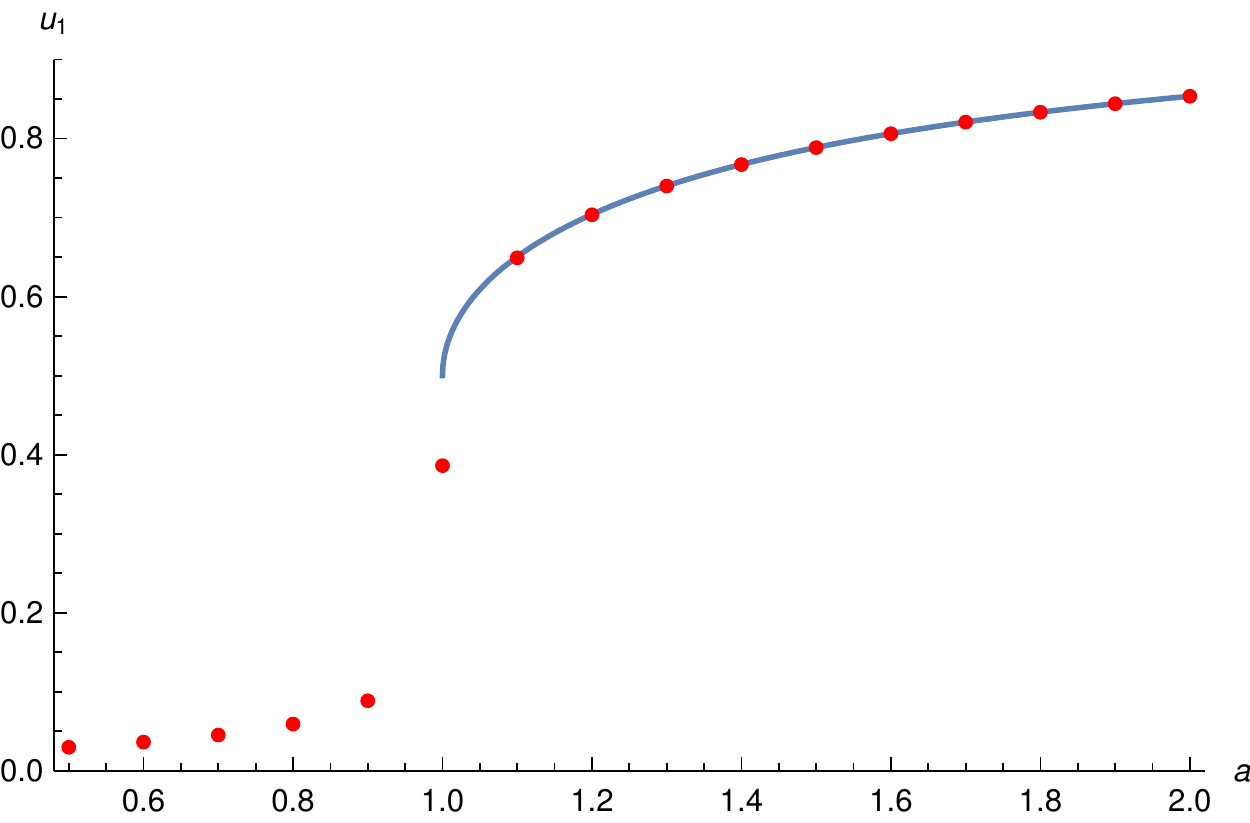}}
\hskip8mm
\subcaptionbox{$u_2$\label{fig:F2}}{\includegraphics[width=4.5cm]{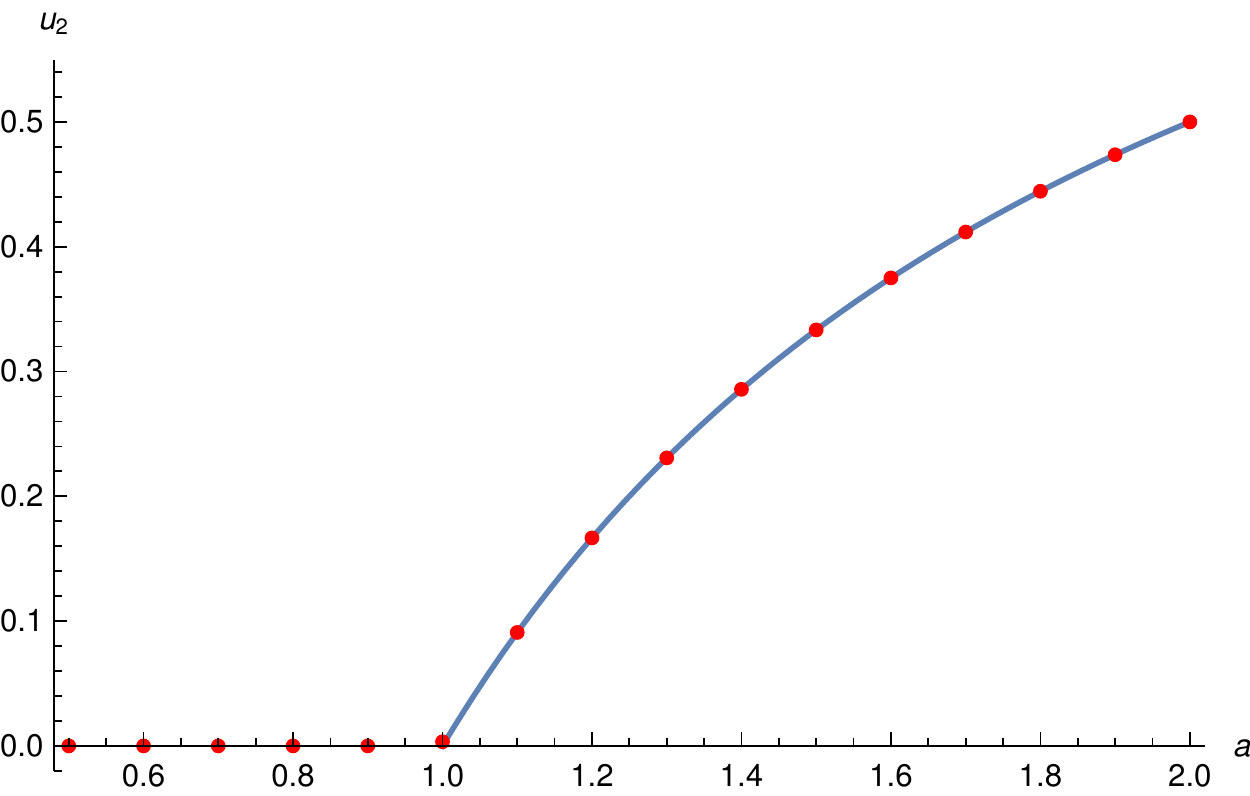}}
\hskip8mm
\subcaptionbox{$u_3$\label{fig:F3}}{\includegraphics[width=4.5cm]{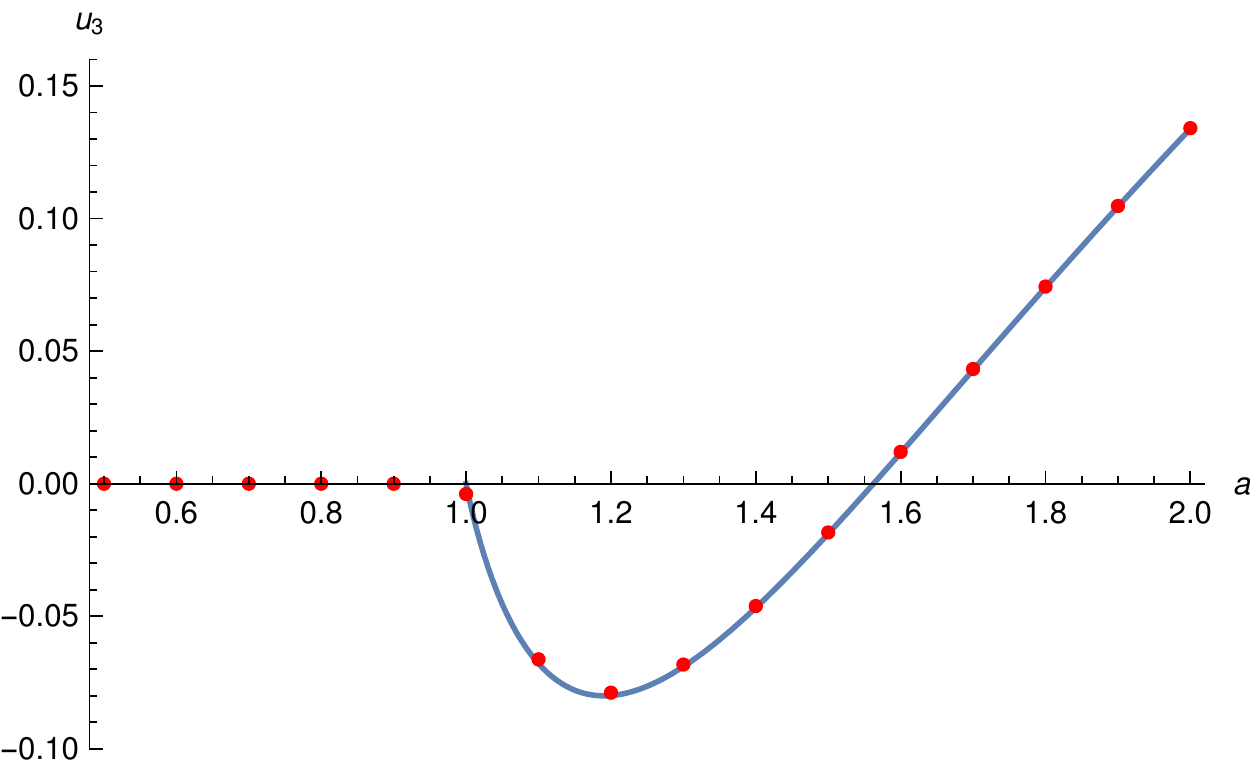}}
  \caption{
Plot 
of the expectation value of the winding Wilson loops 
$u_k=\frac{1}{N}\bra \Tr U^k\ket_a$ in the adjoint model for $k=1,2,3$.
The red dots are the numerical values
at $N=30$, while
solid curves represent the large $N$ result in 
\eqref{eq:uk-saddle}.
}
  \label{fig:a-w123}
\end{figure}

In the case of expectation value of winding loops, 
the integral in the GWW model can be performed in a closed form 
\begin{align}
  \bra \Tr U^k\ket_{a}=
\frac{\int_0^\infty dg g e^{-\frac{N^2g^2}{4a}}\det(M_0)\Tr(M_0^{-1}M_k)}
{\int_0^\infty dg g e^{-\frac{N^2g^2}{4a}}\det(M_0)}.
\label{eq:wind-adj}
\end{align} 
At the leading order in the large $N$ limit, we observed that
the integral over $g$ can be replaced by its saddle point value
\begin{align}
 \frac{1}{N}\bra \Tr U^k\ket_{a}=\frac{1}{N}\bra \Tr U^k\ket\Bigl|_{g=g_*},\qquad (a>1,~N\gg1),
\label{eq:uk-saddle}
\end{align}
where the right-hand-side is the expectation value of
Wilson loop in the GWW model 
evaluated at $g=g_*$.
In Fig.~\ref{fig:a-w123},
we plot the expectation value of winding Wilson loops in the adjoint model.
One can see that the leading result \eqref{eq:uk-saddle}
is reproduced from the numerical evaluation of \eqref{eq:wind-adj} for $N=30$.
As expected, the winding loops are suppressed in the confined phase $a<1$
\begin{align}
 \lim_{N\to\infty} \frac{1}{N}\bra \Tr U^k\ket_{a}=0\qquad(\forall k\geq1),
\end{align}
which is consistent with the absence of non-contractible 1-cycle in the thermal AdS \cite{Witten:1998zw}. It would be interesting to study the (non)perturbative correction to
the winding loops in the large $N$ expansion.

\subsection*{Giant loops in the adjoint model}
\begin{figure}[htb]
\centering
\subcaptionbox{Anti-symmetric representation $W_{A_k}$\label{fig:Ak}}{\includegraphics[width=6cm]{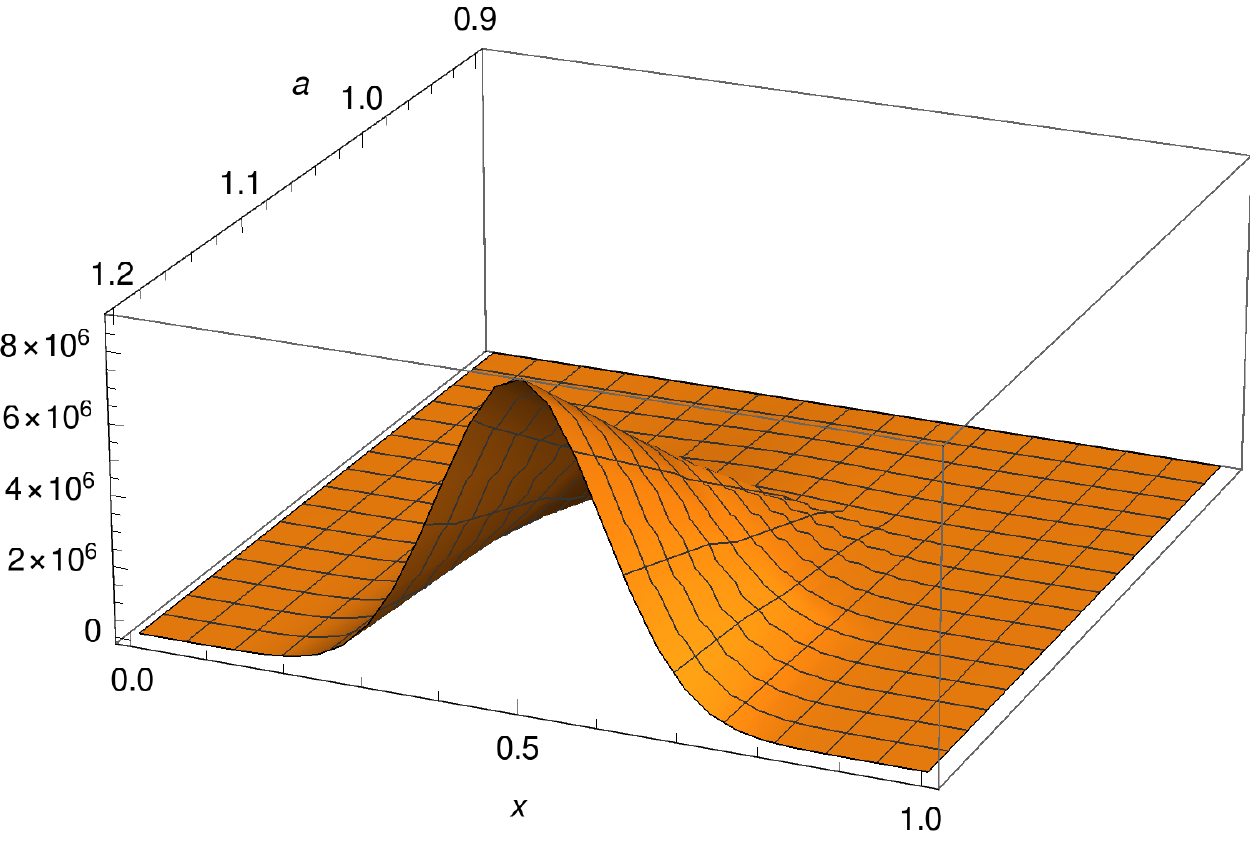}}
\hskip15mm
\subcaptionbox{Symmetric representation $W_{S_k}$\label{fig:Sk}}{\includegraphics[width=6cm]{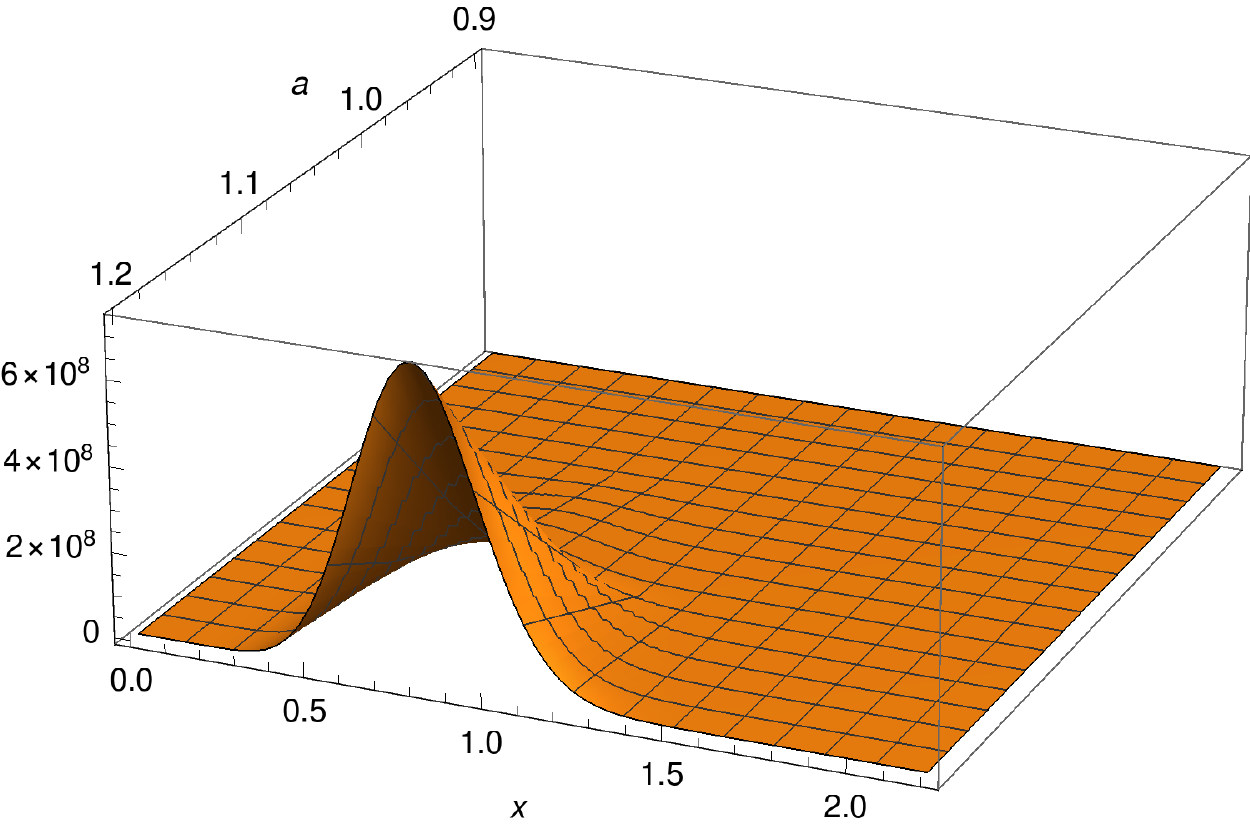}}
  \caption{
Plot 
of the expectation value of Wilson loops in  \subref{fig:Ak}
the anti-symmetric representation and \subref{fig:Sk} the symmetric representation, as functions of 
$a$ and $x=k/N$ for $N=30$. 
}
  \label{fig:a-giant}
\end{figure}

Using the integral transformation \eqref{eq:veva-int}, one can compute
the expectation value of Wilson loops in the adjoint model in arbitrary
representation using the exact result of GWW model
\begin{align}
 \bra \Tr_\la U\ket_{a}=
\frac{\int_0^\infty dg g e^{-\frac{N^2g^2}{4a}}\det\Bigl[I_{\la_j+i-j}(Ng)]}
{\int_0^\infty dg g e^{-\frac{N^2g^2}{4a}}\det\Bigl[I_{i-j}(Ng)\Bigr]} .
\end{align}
In particular, we can study Giant Wilson loops of adjoint model
in the $k$-th (anti)symmetric representation
in the limit \eqref{eq:giant-x}.
At the leading order in the large $N$ limit,
the $g$-integral is approximated by the saddle point value $g=g_*$. 
We have checked numerically that 
the result of \cite{Grignani:2009ua} is reproduced.
As we can see from Fig.~\ref{fig:a-giant},
the expectation values of Giant loops are suppressed in the confined phase $a<1$.
In the deconfined phase,
Giant loop in the symmetric representation
$W_{S_k}$ is exponentially 
suppressed when $x=k/N$ becomes larger than some critical value $x_{\text{cr}}$,
as observed in \cite{Grignani:2009ua}.
It is argued that this is consistent with the absence of D3-brane solution
corresponding to $W_{S_k}$
in the black hole background  \cite{Hartnoll:2006hr,Grignani:2009ua}.
It would be interesting to study the critical value $x_{\text{cr}}$
as a function of $a$ and see if it has some physical interpretation on the dual black hole
side.

\section{Discussion \label{sec:discuss}}
In this paper we have studied the free energy and Wilson loops in the GWW model and the adjoint model
using the exact result at finite $N$.
For the GWW model the exact finite $N$ result correctly reproduces the known
large $N$ expansion of free energy and Wilson loops.
We have also seen that one can extract the (non)perturbative corrections in the large $N$ 
expansion from the exact finite $N$ result by numerical fitting, and some of the results
in this paper are new.
It would be interesting to develop an analytic method to compute such  (non)perturbative corrections
and see if our numerical results are reproduced from analytic computation.

We have seen that the large $N$ expansion of free energy and Wilson loops behaves
quite differently between the gapped phase and the ungapped phase of GWW model.
In the gapped phase the genus expansion is Borel non-summable and the perturbative
and non-perturbative corrections are related by resurgence \cite{Marino:2008ya}.
On the other hand, in the ungapped phase,
the perturbative corrections stop at first order.
Although the instanton coefficient in the ungapped phase has an all order expansion in $1/N$,
this series is Borel summable and the each instanton sector seems to be closed 
by itself (see appendix \ref{app:bessel} for details). This is in stark contrast to the situation in the gapped phase 
and it would be interesting to see how these two expansions are connected when we cross the transition 
point $g=1$.

We proposed a master field of GWW model from the exact result of characteristic polynomial at finite $N$. We found that this master field has an interesting eigenvalue distribution.
In the gapped phase the eigenvalue distribution approaches the known
gapped distribution on the 
unit circle as $N$ becomes large. 
On the other hand, in the ungapped phase we observed that 
the eigenvalues are distributed inside the unit circle 
and we find numerically that 
the eigenvalues are located along the  contour $\Phi(z)=-S_{\text{inst}}(g)$
of constant effective potential.
We do not have a proof of the last statement and it would be interesting to
show this analytically.
Also, it is not clear whether the distribution
on the contour $\Phi(z)=-S_{\text{inst}}(g)$ satisfies the saddle point equation of GWW model or not.
It would be very interesting to clarify the physical interpretation, if any, 
of this distribution further.

We have also studied Giant Wilson loops in both the GWW model and 
the adjoint model. In particular, in the adjoint model Giant Wilson loops are expected to 
be holographically dual to some configuration of D-branes.
We hope that our finite $N$ analysis will shed light on
the behavior of D-branes in black hole background
or the black hole itself
beyond the supergravity approximation.

\vskip8mm
\centerline{\bf Acknowledgments}
\vskip2mm
\noindent
I would like to thank Pavel Buividovich, Gerald Dunne, Marcos Marino, Shunya Mizoguchi, 
Takeshi Morita, Tomoki Nosaka, Semen Valgushev, and Yasuhiko Yamada
for useful discussions and correspondences.
This work  was supported in part by JSPS KAKENHI Grant Number 16K05316.

\appendix
\section{Exact result of GWW model \label{app:exact}}
In this appendix we review the exact result of partition function and Wilson loops in GWW model
at finite $N$.

Let us first consider the following integral with some function $f$
\begin{align}
I^f= \int_{U(N)}dU\det\big[f(U)\big]e^{\frac{Ng}{2}\Tr(U+U^\dagger)}.
\end{align}
This can be rewritten as an integral over the eigenvalues $\{e^{\ri \th_j}\}_{j=1,\cdots,N}$
of unitary matrix $U$
\begin{align}
 I^f=\frac{1}{N!}\int_0^{2\pi}|\lap|^2\prod_{j=1}^N\frac{d\th_j}{2\pi}
e^{Ng\cos\th_j} f(e^{\ri\th_j}),
\label{eq:If-lap}
\end{align}
where $\lap$ denotes the Vandermonde determinant
\begin{align}
 \lap=\sum_{\si\in S_N}(-1)^\si \prod_{j=1}^Ne^{\ri (N-j)\th_{\si(j)}}.
\label{eq:Vdet}
\end{align}
Plugging \eqref{eq:Vdet} into \eqref{eq:If-lap}, we get a double sum over $S_N$.
Since the integrand is symmetric under the permutation of variables $\th_j$,
one can show that this sum can be reduced to a single sum over $S_N$
\begin{align}
 I^f=\sum_{\si\in S_N}(-1)^\si\prod_{j=1}^N\int_0^{2\pi}\frac{d\th_j}{2\pi} 
e^{\ri(\si(j)-j)\th_j}e^{Ng\cos\th_j} f(e^{\ri\th_j})=\det\big[I^f_{i-j}\big]_{i,j=1,\cdots,N},
\end{align} 
where we defined 
\begin{align}
 I_m^f=\int_0^{2\pi}\frac{d\th}{2\pi} 
e^{\ri m\th}e^{Ng\cos\th} f(e^{\ri\th}).
\label{eq:If-moment}
\end{align}

For the computation of partition function, we set $f=1$.
Then the integral  \eqref{eq:If-moment} is nothing but the modified Bessel function 
of the first kind $I_m(Ng)$, and we recover the exact result of partition function at finite $N$
in \eqref{eq:Z-exact}.

For the computation of winding Wilson loop $\Tr U^k$, we set
\begin{align}
 f(U)=1+t U^k
\end{align}
and pick up the linear term of $t$ in the small $t$ expansion
\begin{align}
 \det\big[f(U)\big]=1+t\Tr U^k+\mathcal{O}(t^2).
\end{align}
For this choice of $f$, the integral $I_m^f$ in \eqref{eq:If-moment} becomes
\begin{align}
 I^f_m=I_m(Ng)+tI_{k+m}(Ng),
\end{align}
and we find
\begin{align}
 I^f=\det\Big[I_{i-j}(Ng)+tI_{k+i-j}(Ng)\Big]=\det(M_0+t M_k).
\label{eq:If-wind}
\end{align}
Here the $N\times N$ matrix $M_k$ has been defined in \eqref{eq:Mk}.
Picking up the linear term in $t$ and normalizing 
by the partition function $Z(N,g)=\det M_0$, we find that 
the expectation value of winding Wilson loop $\bra \Tr U^k\ket$ is given by \eqref{eq:exact-tr}.
In a similar manner, one can show the relation \eqref{eq:exact-ch}
\begin{align}
\begin{aligned}
 \bra \det(x-U)\ket&=\frac{1}{Z(N,g)}\int dU\det(x-U)e^{\frac{Ng}{2}\Tr(U+U^\dagger)}\\
&=\frac{\det(xM_0-M_1)}{\det M_0}=\det(x-M_0^{-1}M_1). 
\end{aligned}
\end{align}

Lastly, let us consider the expectation value of the character $\Tr_\la U=\chi_\la(U)$ of $U(N)$
group
\begin{align}
 \chi_\la(U)=\frac{1}{\lap}\sum_{\si\in S_N}(-1)^\si \prod_{j=1}^N e^{\ri (\la_j+N-j)\th_{\si(j)}}.
\end{align}
Again, the factor $|\lap|^2 \chi_\la(U)$ becomes a double sum over the permutation group $S_N$,
but this sum can be reduced to a single sum upon integration
and we find
\begin{align}
 \int dU \chi_\la(U)e^{\frac{Ng}{2}\Tr (U+U^\dagger)}=\sum_{\si\in S_N}
(-1)^\si \prod_{j=1}^N \int_0^{2\pi}\frac{d\th_j}{2\pi} e^{\ri (\la_j-j+\si(j))\th_{j}}e^{Ng\cos\th_j}
=\det\Big[I_{\la_j+i-j}(Ng)\Big].
\end{align}
After dividing by the partition function, we recover the result 
of $\bra \Tr_\la U\ket$ in \eqref{eq:Trla-exp}.

\section{Effective potential in the ungapped phase \label{app:potential}}
In this appendix, we explain the computation of the effective potential $\Phi(z)$ in \eqref{eq:Phi-z}
following the argument in \cite{Alvarez:2016rmo}. 
As discussed in \cite{Alvarez:2016rmo}, the eigenvalue integral \eqref{eq:If-lap}
can be rewritten as a holomorphic integral
with complex variable $z_j=e^{\ri\th_j}$.
For the partition function we find
\begin{align}
 Z(N,g)=\frac{1}{N!}\int \prod_{j=1}^N\frac{dz_j}{2\pi\ri }e^{
-NW(z_j)}
\prod_{i<j}(z_i-z_j)^2,
\label{eq:hol-model}
\end{align}
where the potential $W(z)$ is given by
\begin{align}
 W(z)=-\frac{g}{2}(z+z^{-1})+\log z.
\end{align}
The integral \eqref{eq:hol-model}
has the same form as the hermitian matrix model, although the integral contour
is different: in the unitary matrix model the integral contour is along the unit circle
$|z_j|=1$ while in the hermitian matrix model the integral is along the real axis
$z_j\in\mathbb{R}$.
At least formally, the saddle point equation for the eigenvalue integral \eqref{eq:hol-model}
takes the same form as that of the hermitian matrix model
\begin{align}
 W'(z_i)-\frac{2}{N}\sum_{j\not=i}\frac{1}{z_i-z_j}=0.
\end{align}
Then one can show that the resolvent defined by
\begin{align}
 \om(z)=\frac{1}{N}\sum_{i=1}^N\frac{1}{z-z_i}
\end{align}
satisfies the loop equation
\begin{align}
 \om(z)^2+\frac{1}{N}\om'(z)-W'(z)\om(z)+f(z)=0,
\label{eq:loop}
\end{align}
where $f(z)$ is given by
\begin{align}
 f(z)=\frac{1}{N}\sum_{i=1}^N\frac{W'(z)-W'(z_i)}{z-z_i}.
\end{align}
In the planar limit, the second term of \eqref{eq:loop} can be omitted and 
the loop equation can be written as
an algebraic equation defining a spectral curve
\begin{align}
 y^2=W'(z)^2-4f(z)
\label{eq:spec}
\end{align}
with $y$ being
\begin{align}
 y=W'(z)-2\om(z).
\label{eq:ydef}
\end{align} 
As emphasized in \cite{Dijkgraaf:2002fc}, the quantity $y$ has an elegant physical
interpretation as the force acting on an eigenvalue
if it tries to move away from its stationary position.
This suggests that it is natural to define an effective potential
as the integral of force: $U(z)=\int^z ydz$.
However, as discussed in  \cite{Alvarez:2016rmo},
it is more appropriate to take the real part of $\int^z ydz$ and define
the effective potential as
\begin{align}
 \Phi(z)=\text{Re}\int^z ydz,
\label{eq:Phi-ydz}
\end{align}
since the dominance to the eigenvalue integral \eqref{eq:hol-model}
is dictated by the real part of potential.
One can show that the potential $\Phi(z)$ is constant on each cut made by the
condensation of eigenvalues in the large $N$ limit.

Now let us compute the effective potential in the ungapped phase of GWW model.
To do this, we notice that the planar resolvent in the gapped phase has a simple expansion in the 
large $z$ region
\begin{align}
\om(z)=\frac{1}{z}+\sum_{k=1}^\infty \frac{1}{N}\bra \Tr U^k\ket \frac{1}{z^{k+1}}=
\frac{1}{z}+\frac{g}{2z^2},	
\end{align}
since winding Wilson loops $\bra\Tr U^k\ket$ vanish except for $k=1$
(see \eqref{eq:u1-vev} and \eqref{eq:uk-lead}).
Then the quantity $y$ in \eqref{eq:ydef} is given by
\begin{align}
 y=-\frac{g}{2}\Bigl(1+\frac{1}{z^2}\Bigr)-\frac{1}{z},
\end{align}
and the spectral curve \eqref{eq:spec} becomes
\begin{align}
 y^2=\left[\frac{g}{2}\Bigl(1+\frac{1}{z^2}\Bigr)+\frac{1}{z}\right]^2.
\end{align}
This curve has two branches and we should be careful about the  sign of $y$.
Assuming that the eigenvalues are distributed along the unit circle $|z|=1$,
 the sign of $y$ should change as we cross the line $|z|=1$
\begin{align}
 y=\left\{
\begin{aligned}
-\frac{g}{2}\Bigl(1+\frac{1}{z^2}\Bigr)-\frac{1}{z} ,&\qquad (|z|>1),\\
+\frac{g}{2}\Bigl(1+\frac{1}{z^2}\Bigr)+\frac{1}{z},&\qquad (|z|<1),
\end{aligned}
\right.
\end{align}
One can show that the eigenvalue density $\rho(\th)$ \eqref{eq:rho}
is reproduced from the discontinuity along $|z|=1$.
Finally, the effective potential $\Phi(z)$ is given by 
the integral \eqref{eq:Phi-ydz}
and we arrive at the result \eqref{eq:Phi-z}.

\section{Instanton correction in the ungapped phase \label{app:bessel}}
In this appendix, we consider
the instanton correction of free energy in the ungapped phase of GWW model.
Here (and only in this appendix)
we use the convention of string coupling $g_s$  and 't Hooft coupling $t$
in footnote \ref{foot:Z-gs}:  
\begin{align}
 Z(N,g_s)=\int_{U(N)} dU \exp\left[\frac{1}{2g_s}\Tr(U+U^\dagger)\right]=
\det\Bigl[I_{i-j}(1/g_s)\Bigr]_{i,j=1,\cdots,N}.
\end{align}
We are interested in the instanton corrections in the 't Hooft limit
\begin{align}
 N\to\infty,~~ g_s\to0,\qquad t=Ng_s:~\text{fixed}.
\label{eq:thooft-lim}
\end{align}
Instanton corrections to the free energy in the gapped phase $t<1$
have been studied extensively in \cite{Marino:2008ya}. 
Here we would like to point out that the first non-zero
instanton correction to the free energy in the ungapped phase $t>1$
can be written in a closed form.

To study the (non)perturbative corrections to the free energy, it is convenient 
to use the method of orthogonal polynomial
$p_n(z)$ obeying
\begin{align}
 \oint\frac{dz}{2\pi\ri z}e^{\frac{1}{2g_s}(z+z^{-1})}p_n(z)p_m(z^{-1})=h_n\cob_{n,m}.
\end{align}
The partition function of  GWW model is written in terms of the norm $h_n$ as
\begin{align}
 Z(N,g_s)=\prod_{n=0}^{N-1}h_n.
\label{eq:Z-hn}
\end{align}
From the constant term $f_n$ of $p_n(z)$ \footnote{Note that we have shifted the index
$n$ of $f_n$ by one as compared to the definition of \cite{Marino:2008ya}.}
\begin{align}
 f_n=(-1)^np_n(0)
\label{eq:fn-pn}
\end{align}
we can compute the ratio of the norm $h_n$
\begin{align}
 \frac{h_n}{h_{n-1}}=1-f_n^2.
\label{eq:hn-vs-fn}
\end{align}
From \eqref{eq:Z-hn} and \eqref{eq:hn-vs-fn}
one can show that
\begin{align}
 \frac{Z(N+1,g_s)Z(N-1,g_s)}{Z(N,g_s)^2}=1-f_N^2.
\label{eq:ZN-rec}
\end{align}
Furthermore, using  the recursion relation
\begin{align}
 p_{n}(z)=zp_{n-1}(z)+(-1)^n f_n z^{n-1}p_{n-1}(z^{-1}),
\end{align}
one can show that $f_n$ satisfies
\begin{align}
 2g_sn f_n=(1-f_n^2)(f_{n+1}+f_{n-1}).
\label{eq:fn-rec}
\end{align}
Note that this is known as a discrete  Painlev\'{e} equation \cite{Hisakado:1996di,TW}.
From Heine's formula 
the orthogonal polynomial $p_n(z)$ with $n=N$
is simply given by the expectation
value of the characteristic polynomial in the GWW model
\begin{align}
 p_N(z)=\bra \det(z-U)\ket.
\end{align}
This also implies that $f_n$ \eqref{eq:fn-pn} with $n=N$
is given by
the expectation value of $\det U$
\begin{align}
 f_N=\bra \det U\ket=\frac{\det\Bigl[I_{1+i-j}(1/g_s)\Bigr]}{\det\Bigl[I_{i-j}(1/g_s)\Bigr]}.
\end{align}

In the 't Hooft limit \eqref{eq:thooft-lim}, $f_N$ becomes a function $f(t,g_s)$ of 
the 't Hooft coupling $t$ and the string coupling $g_s$.
Then $f(t,g_s)$ satisfies the continuum version of the recursion relation \eqref{eq:fn-rec}
\begin{align}
 2t f(t,g_s)=\Bigl(1-f(t,g_s)^2\Bigr)\Bigl(f(t+g_s,g_s)+f(t-g_s,g_s)\Bigr).
\label{eq:f-rec}
\end{align}
This is called the {\it pre-string} equation.
Once we know the function $f(t,g_s)$, we can compute the free energy $F(t,g_s)$ from
the continuum limit of \eqref{eq:ZN-rec}
\begin{align}
 F(t,g_s)=\frac{1}{4\sinh^2\frac{g_s\del_t}{2}}\log R(t,g_s),
\label{eq:F-to-R}
\end{align}
where $R(t,g_s)$ is defined by
\begin{align}
 R(t,g_s)=1-f(t,g_s)^2.
\label{eq:Rdef}
\end{align}

In the ungapped phase, $f(t,g_s)$ is exponentially small. Thus 
the  relation \eqref{eq:f-rec} is approximated by
\begin{align}
 2t f^{(1)}(t,g_s)=f^{(1)}(t+g_s,g_s)+f^{(1)}(t-g_s,g_s),
\end{align}
where we  have introduced the notation $f^{(1)}(t,g_s)$ for the one-instanton correction
to $f(t,g_s)$.
We notice that this is exactly the recursion relation of Bessel function $J_\nu(x)$
\begin{align}
 2Ng_s J_N(1/g_s)=J_{N+1}(1/g_s)+J_{N-1}(1/g_s).
\end{align}
Thus we expect that $f(t,g_s)$ is proportional to $J_{N}(1/g_s)=J_{N}(N/t)$,
which is consistent with 
the large $N$ behavior of $\bra \det U\ket$  studied in \cite{Rossi:1982vw}.

As discussed in \cite{Marino:2008ya}, we can fix the proportionality constant by
comparing the double-scaling limit of $J_{N}(N/t)$
and the Hastings-McLeod solution of the Painlev\'{e} II equation.
In the double scaling limit
\begin{align}
 f=g_s^{1/3}u,\quad t=1-g_s^{2/3}\ka,\quad g_s\to0,
\end{align}
$u(\ka)$ satisfies the Painlev\'{e} II equation
\begin{align}
 u''-2u^3+2\ka u=0.
\end{align}
There is a unique real solution (Hastings-McLeod solution) for $\ka\in\mathbb{R}$
with the asymptotic behavior
\begin{align}
 u=
 \left\{
  \begin{aligned}
&\rt{\ka},&\qquad &(\ka\to\infty),\\
&2^{\frac{1}{3}}\text{Ai}(-2^{\frac{1}{3}}\ka),
&\qquad &(\ka\to-\infty).\\
 \end{aligned}
\right.
\label{eq:asy-uHM}
\end{align}
One can compare this with the double scaling limit 
of the Bessel function \cite{dlmf1}
\begin{align}
 \lim_{N\to\infty }N^{\frac{1}{3}}J_N(N+N^{\frac{1}{3}}\ka)=2^{\frac{1}{3}}\text{Ai}(-2^{\frac{1}{3}}\ka).
\label{eq:dsl-bessel}
\end{align}
From \eqref{eq:asy-uHM} and \eqref{eq:dsl-bessel}, we conclude that
the proportionality constant is 1
\begin{align}
 f^{(1)}(t,g_s)=J_{N}(N/t),\qquad N=\frac{t}{g_s}.
\label{eq:finst}
\end{align}

Now we can study the genus expansion of 1-instanton coefficients
in the ungapped phase $(t>1)$
using the so-called Debye expansion of Bessel function \cite{dlmf2}
\begin{align}
 J_N(N/\cosh\al)=\frac{e^{-N(\al-\tanh\al)}}{\rt{2\pi N \tanh\al}}
\sum_{k=0}^\infty \frac{U_k(\coth\al)}{N^k},
\label{eq:debye}
\end{align}
where $U_k(x)$ is a polynomial defined recursively from $U_0=1$
\begin{align}
U_{k+1}(x)=\hf x^2(1-x^2)U'_k(x)+\frac{1}{8}\int_0^xdy (1-5y^2)U_k(y).
\end{align}
The first three terms are given by
\begin{align}
\begin{aligned}
 U_1(x)&=\frac{-5 x^3+3 x}{24},\\
U_2(x)&=\frac{385 x^6-462 x^4+81 x^2}{1152},\\
U_3(x)&=\frac{-425425 x^9+765765 x^7-369603 x^5+30375
   x^3}{414720} .
\end{aligned}
\end{align}

From \eqref{eq:finst} and \eqref{eq:debye}, we identify $\cosh\al=t$.
Finally we arrive at a closed form of 1-instanton correction in the ungapped phase
\begin{align}
 f^{(1)}(t,g_s)=
\rt{\frac{g_s}{2\pi}}\frac{e^{-\frac{1}{g_s}A(t)}}{(t^2-1)^{1/4}}\sum_{k=0}^\infty g_s^k t^{-k}U_k\Biggl(\frac{t}{\rt{t^2-1}}\Biggr),
\label{eq:f1-sum}
\end{align}
where the instanton action $A(t)$ is given by
\begin{align}
 A(t)=t(\al-\tanh\al)=t\cosh^{-1}(t)-\rt{t^2-1}.
\end{align}
From the relation \eqref{eq:Rdef}
the two-instanton correction to $R(t,g_s)$ is given by
\begin{align}
 \begin{aligned}
  R^{(2)}(t,g_s)&=-f^{(1)}(t,g_s)^2\\
&=-\frac{g_s}{2\pi}\frac{e^{-\frac{2}{g_s}A(t)}}{\rt{t^2-1}}
\Biggl[\sum_{k=0}^\infty g_s^k t^{-k}U_k\Biggl(\frac{t}{\rt{t^2-1}}\Biggr)\Biggr]^2\\
&=-\frac{g_s}{2\pi}\frac{e^{-\frac{2}{g_s}A(t)}}{\rt{t^2-1}}\left[1-g_s\frac{2 t^2+3}{12 \left(t^2-1\right)^{3/2}}+g_s^2
\frac{4 t^4+156 t^2+45}{288 \left(t^2-1\right)^3}+\cdots\right].
 \end{aligned}
\label{eq:Rtwo}
\end{align}
This agrees with the result of \cite{Marino:2008ya} obtained by solving the {\it pre-string}
equation \eqref{eq:f-rec}, but the overall factor was not determined in \cite{Marino:2008ya}. 
We have fixed the overall normalization
of $R^{(2)}(t,g_s)$ as discussed above.
Now the result \eqref{eq:Rtwo}
can be easily translated to the two-instanton correction
to the free energy using the relation \eqref{eq:F-to-R}
\begin{align}
 F^{(\text{2-inst})}=-\frac{\h{g}_s}{8\pi}e^{-\frac{2}{g_s}A(t)}
\Biggl[1-\h{g}_s\frac{26 t^2+9}{12}+\h{g}_s^2\frac{964 t^4+2484 t^2+297}{288}+\cdots\Biggr],
\end{align}
where we have introduced the rescaled coupling $\h{g}_s$ by
\begin{align}
 \h{g}_s=\frac{g_s}{\left(t^2-1\right)^{3/2}}.
\end{align}

\begin{figure}[tbh]
\centering\includegraphics[width=8cm]{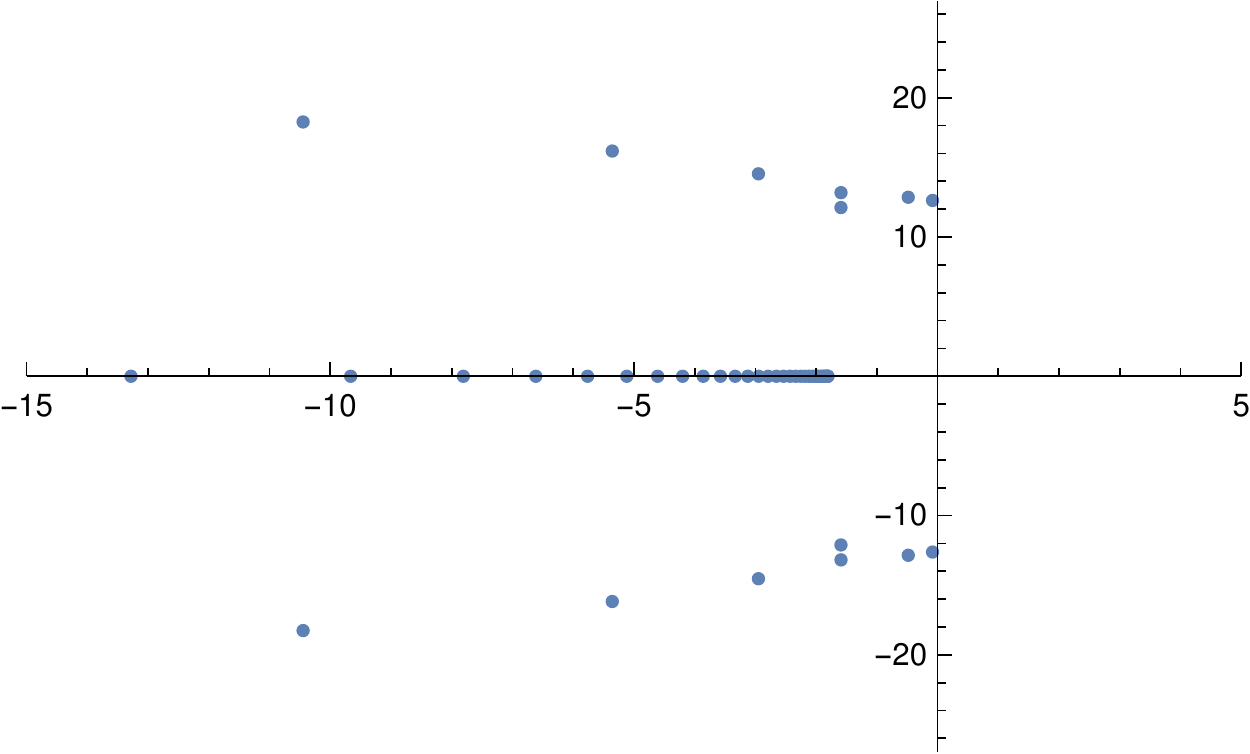}
  \caption{
Poles of the integrand of \eqref{eq:borel-sum} 
on the Borel $\zeta$-plane for $t=2$. 
}
  \label{fig:borel}
\end{figure}

It is interesting to consider the Borel summability of the Debye expansion in \eqref{eq:f1-sum}.
Let us consider the Borel sum
\begin{align}
 \mathcal{B}\left[\sum_{k=0}^\infty g_s^k t^{-k}U_k\Biggl(\frac{t}{\rt{t^2-1}}\Biggr)\right]
=\int_0^\infty\frac{d\zeta}{g_s}e^{-\frac{\zeta}{g_s}}\sum_{k=0}^\infty 
\frac{\zeta^k}{k!} t^{-k}U_k\Biggl(\frac{t}{\rt{t^2-1}}\Biggr).
\label{eq:borel-sum}
\end{align}
As we can see from Fig. \ref{fig:borel}, there is no pole on the positive real axis
on the Borel plane and hence the expansion of $f^{(1)}(t,g_s)$ in \eqref{eq:f1-sum}
is Borel summable. We have checked numerically
that the Borel resummation of $f^{(1)}(t,g_s)$
agrees with the original expression of Bessel function \eqref{eq:finst}.
This is in a stark contrast to the situation in the gapped phase.
As shown in \cite{Marino:2008ya}, in the gapped phase the genus expansion of free energy is
Borel non-summable and the perturbative part and 
the non-perturbative part are related by the resurgence.
On the other hand, in the ungapped phase 
the perturbative genus expansion of free energy
is not an infinite power series but stops at genus-zero.
Although the one-instanton coefficient has infinite series expansion
in $g_s$, it is Borel summable as we have seen above.

\section{Resolvent of GWW model \label{app:resolvent}}
In this appendix we consider the genus-one resolvent of GWW model in the gapped phase, 
from which
we can extract the genus-one correction to the winding Wilson loops
and compare with the result of numerical fitting \eqref{eq:trU-genus}.
To do this, we use the relation between unitary matrix model and hermitian matrix model \cite{Mizoguchi:2004ne} and the formula of genus-one resolvent of hermitian matrix model \cite{Ambjorn:1992gw}.

As shown in \cite{Mizoguchi:2004ne}, a unitary matrix model
can be written as a hermitian matrix model
\begin{align}
 \int dU e^{-N\Tr V(U)}=\int dM e^{-N\Tr W(M)}
\label{eq:U-vs-H}
\end{align}
where the eigenvalue $t$ of unitary matrix $U$
and the eigenvalue $z$ of hermitian matrix $M$ are related by 
\begin{align}
 t=\frac{1+\ri z}{1-\ri z},
\end{align}
and the potentials in \eqref{eq:U-vs-H}
are related by
\begin{align}
 W(z)=V(t)+\log(1+z^2).
\end{align}

In the case of GWW model the potential are given by
\begin{align}
 \begin{aligned}
  V(t)&=-\frac{g}{2}(t+t^{-1}),\qquad
W(z)=g\frac{z^2-1}{z^2+1}+\log(1+z^2).
 \end{aligned}
\end{align}
We define the resolvent $\om(z)$ of hermitian matrix model
and the resolvent $v(t)$ of unitary matrix model as
\begin{align}
\begin{aligned}
 \om(z)&=\frac{1}{N}\left\bra \Tr\frac{1}{z-M}\right\ket,\\
v(t)&=\frac{\ri}{N}\left\bra \Tr\frac{t+U}{t-U}\right\ket,
\end{aligned}
\end{align}
and they are related by
\begin{align}
 v(t)=(1+z^2)\om(z)-z.
\label{eq:res-dict}
\end{align}
In the large $N$ limit these resolvents have genus expansion
\begin{align}
 \begin{aligned}
  \om(z)=\sum_{\ell=0}^\infty N^{-2\ell}\om_\ell(z),\qquad
v(t)=\sum_{\ell=0}^\infty N^{-2\ell}v_\ell(t).
 \end{aligned}
\end{align}
Using the technique developed in \cite{Ambjorn:1992gw}
for hermitian matrix model,
one can compute the higher genus correction
of resolvent $\om_\ell(z)$ recursively.
In what follows we assume that the hermitian matrix model is in the one-cut phase, i.e.
eigenvalues
are distributed along the cut $z\in[-A,A]$ 
on the real axis.

\subsection*{Genus-zero resolvent}
Let us first consider the genus-zero resolvent which is given by
the general formula
\begin{align}
 \om_0(z)=
\int_C\frac{dx}{4\pi\ri}\frac{W'(x)}{z-x}\rt{\frac{z^2-A^2}{x^2-A^2}},
\label{eq:om0}
\end{align}
where the contour $C$ encircles the cut $[-A,A]$.
From the condition 
\begin{align}
 \lim_{z\to \infty}\om_0(z)=\frac{1}{z}+\mathcal{O}(z^{-2})
\end{align}
we find
\begin{align}
 \int_C\frac{dx}{4\pi\ri}\frac{W'(x)}{\rt{x^2-A^2}}=0,\qquad
\int_C\frac{dx}{4\pi\ri}\frac{xW'(x)}{\rt{x^2-A^2}}=1.
\end{align}
From these conditions we can fix
the end-point of cut $A$ as a function of coupling $g$
\begin{align}
 A=\frac{1}{\rt{g-1}}.
\end{align}
Picking up the residue of poles at $x=\pm\ri$ and $x=\infty$ in \eqref{eq:om0},
the genus-zero resolvent becomes
\begin{align}
 \om_0(z)=\hf\left[W'(z)-M(z)\rt{z^2-A^2}\right],
\qquad
M(z)=\frac{4\rt{1+A^2}}{A^2(1+z^2)^2}.
\label{eq:Mz}
\end{align}
Then using the dictionary between resolvents of hermitian and 
unitary matrix models \eqref{eq:res-dict},
we arrive at the genus-zero resolvent of GWW model
\begin{align}
 v_0(t)=\frac{2g}{1+z^2}\left[z-\rt{z^2-g^{-1}(1+z^2)}\right].
\end{align}
We note in passing that one can easily show that 
this agrees with the integral over the eigenvalues $e^{\ri\th}$ 
with the weight $\rho(\th)$ 
in the gapped phase \eqref{eq:rho} 
\begin{align}
 \frac{\ri}{2}v_0(t)=\hf \int d\th\rho(\th)\frac{1+te^{\ri\th}}{1-te^{\ri\th}}
=\frac{g(t+1)}{4t}\left[t-1+\rt{(t-1)^2+\frac{4t}{g}}\right].
\end{align}

\subsection*{Genus-one resolvent}
Let us move on to the genus-one resolvent.
The genus-one resolvent in the one-cut phase of hermitian matrix model
is given by \cite{Ambjorn:1992gw}
\begin{align}
 \om_1(z)=\frac{\chi_{+}^{(2)}+\chi_{-}^{(2)}}{16}-\frac{\chi_{+}^{(1)}-\chi_{-}^{(1)}}{16A},
\label{eq:res1}
\end{align}
where $\chi_{\pm}^{(1)}$  and $\chi_{\pm}^{(2)}$ are defined by
\begin{align}
 \begin{aligned}
  \chi_{\pm}^{(1)}&=\frac{1}{M_1\rt{z^2-A^2}(z\mp A)},\\
\chi_{\pm}^{(2)}&=\frac{1}{M_1\rt{z^2-A^2}(z\mp A)^2}\mp\frac{M_2 \chi_{\pm}^{(1)}}{M_1},
 \end{aligned}
\label{eq:chi}
\end{align}
and the moment $M_{k}$ is defined by
\begin{align}
  M_k=\int_C\frac{dx}{2\pi\ri}\frac{W'(x)}{(x-A)^k\rt{x^2-A^2}}=
\frac{1}{(k-1)!}\frac{d^{k-1}}{dz^{k-1}}M(z)\Bigl|_{z=A}.
\end{align}
From the explicit form of function $M(z)$ in \eqref{eq:Mz},
the moments are evaluated as
\begin{align}
 M_1=\frac{4}{A^2(1+A^2)^{\frac{3}{2}}},\qquad
M_2=-\frac{16}{A(1+A^2)^{\frac{5}{2}}}.
\label{eq:moment}
\end{align}
Plugging \eqref{eq:chi} and \eqref{eq:moment} into 
\eqref{eq:res1}, we find the closed form of genus-one resolvent
\begin{align}
 \om_1(z)=\frac{A^4\rt{1+A^2}(2z^2+1-A^2)}{16(z^2-A^2)^{\frac{5}{2}}}.
\end{align}
We can translated this result to the unitary GWW model 
using the dictionary \eqref{eq:res-dict}
\begin{align}
 \begin{aligned}
  \frac{\ri}{2}v_1(t)&=\frac{\ri}{2}(1+z^2)\om_1(z)\\
&=\frac{t(t+1)}{8(g-1)g^2}\left[-g\left((t-1)^2+\frac{4t}{g}\right)^{-\frac{3}{2}}
+4t(g-1)\left((t-1)^2+\frac{4t}{g}\right)^{-\frac{5}{2}}\right].
 \end{aligned}
\end{align}
Finally, we can see that the small $t$ expansion of $v_1(t)$
reproduces the genus-one part of winding
Wilson loops in \eqref{eq:trU-genus}
\begin{align}
\begin{aligned}
 \frac{\ri}{2}v_1(t)
&=-\frac{t}{8(g-1)g}+\frac{t^2}{4(g-1)g^2}+\frac{(10-28g+15g^2)t^3}{8(g-1)g^3}\\
&+
\frac{(-35+90g-70g^2+16g^3)t^4}{2(g-1)g^4}
+\frac{5 \left(35 g^4-260 g^3+630 g^2-616 g+210\right)
   t^5}{8 (g-1) g^5}\\
&+
\frac{\left(192 g^5-2135 g^4+8120 g^3-13860 g^2+10920
   g-3234\right) t^6}{4 (g-1) g^6}+\cdots. 
\end{aligned}
\end{align}

 
\end{document}